\documentclass[useAMS,usenatbib]{mn2e}
\usepackage{graphicx}

\newcommand{\be}{\begin{equation}}
\newcommand{\ee}{\end{equation}}
\newcommand{\bea}{\begin{eqnarray}}
\newcommand{\eea}{\end{eqnarray}}

\title[NoSOCS in SDSS. I]{NoSOCS in SDSS. I. Sample Definition and Comparison of Mass Estimates}
\author[Lopes et al.]{P. A. A. Lopes$^{1}$\thanks{E-mail: 
paal@univap.br}, R. R. de Carvalho$^{2}$, J. L. Kohl-Moreira$^{3}$,
C. Jones$^{4}$\\
$^{1}$IP\&D Universidade do Vale do Para{\' \i}ba, 
Av. Shishima Hifumi 2911, \\ 
S\~ao Jos\'e dos Campos, SP 12244-000, Brasil \\ 
$^{2}$Instituto Nacional de Pesquisas Espaciais -- Divis\~ao de
Astrof{\' \i}sica (CEA), Avenida dos Astronautas, 1758\\
S\~ao Jos\'e dos Campos, SP 12227-010, Brasil \\
$^{3}$Observat\'orio Nacional/MCT, COAA, Brasil \\
$^{4}$Harvard-Smithsonian Center for
Astrophysics, 60 Garden Street, Cambridge, MA 02138, USA}
\begin{document}

\date{Accepted  Received ; in original form }

\pagerange{\pageref{firstpage}--\pageref{lastpage}} \pubyear{2007}

\maketitle

\label{firstpage}

\begin{abstract}

We use Sloan Digital Sky Survey (SDSS) data to investigate galaxy cluster properties of systems first detected within DPOSS. With the high quality photometry of SDSS we derived new photometric redshifts and estimated richness and optical luminosity. For a subset of low redshift ($z \le 0.1$) clusters, we have used SDSS spectroscopic data to identify groups in redshift space in the region of each cluster, complemented with massive systems from the literature to assure the continuous mass sampling. A method to remove interlopers is applied, and a virial analysis is performed resulting in estimates of velocity dispersion, mass, and a physical radius for each low-$z$ system. We discuss the choice of maximum radius and luminosity range in the dynamical analysis, showing that a spectroscopic survey must be complete to at least M$^*+1$ if one wishes to obtain accurate and unbiased estimates of velocity dispersion and mass. We have measured X-ray luminosity for all clusters using archival data from RASS. For a smaller subset (twenty-one clusters) we selected temperature measures from the literature and estimated mass from the M-T$_X$ relation, finding that they show good agreement with the virial estimate. However, these two mass estimates tend to disagree with the caustic results. We measured the presence of substructure in all clusters of the sample and found that clusters with substructure have virial masses higher than those derived from T$_X$. This trend is not seen when comparing the caustic and X-ray masses. That happens because the caustic mass is estimated directly from the mass profile, so it is less affected by substructure.

\end{abstract}


\section{Introduction}

Galaxy clusters represent the largest and latest systems to form under the influence of their own gravity. The study of the evolution of the large scale structure through the cluster mass function and how that varies with time, is one of the most important tools in unveiling the global properties of the Universe. The way clusters evolve strongly depend on several cosmological parameters like $\Omega_m$, $\Omega_{\lambda}$, and $\sigma_8$ \citep {eke98, bah97, bah03, don98, rei99, bla98}.  The main difficulty encountered in these studies, lies in the exact estimate of the cluster mass. A variety of techniques are in use today, but accurate estimates for large data sets are still impractical, as all methods are expensive from the observational viewpoint. 

A simple way to estimate mass for a large sample of clusters is to find an unbiased photometric proxy such as X-ray luminosity (L$_X$), richness (N$_{gal}$), or optical luminosity (L$_{opt}$). L$_X$ and L$_{opt}$ correlate well with mass, though with larger scatter than relations involving X-ray temperature (T$_X$) and velocity dispersions ($\sigma_P$). The measurement and calibration of these relations are essential for reliably establishing the cluster mass function. As pointed out by \citet {voi05}, underestimation of the scatter in the mass$-$observable relation leads to overestimation of $\sigma_8$. Furthermore, contradictory results appear when one relates mass to cluster properties measured at different wavelengths. This maybe reflecting  the selection function of the cluster catalogs created from data in different wavelength regimes, or the physical processes affecting the cluster galaxy population and the intra-cluster gas.

The study of cosmology using galaxy clusters has as a key ingredient the proper recognition of the biases present in multi-wavelength surveys. Some of these issues have been addressed in the recent literature, either by comparing X-ray and optical cluster catalogs or conducting joint X-ray/optical surveys of galaxy clusters \citep {don01, don02, gil04, pop04, pop05, lop06, dai07, gal08}. Some of these works investigate the complementary properties of clusters selected in other wavelength regimes (optical or X-ray; \citealt {don02, gil04, pop04, lop06, dai07, gal08}).

Information on the distribution of galaxies, hot gas and dark matter within clusters can play an important role to understand which observational biases contribute to increasing the scatter of the scaling relations. The presence of substructure is a clear sign of incomplete relaxation in a cluster and can be seen from the X-ray emission from the intra-cluster medium \citep{jon99} as well as in the distribution of cluster galaxies \citep {pin96, mat99, kol01}. Recent works have investigated the possibility that substructures in clusters increase the scatter of the scaling relations, but the results have been contradictory. \citet {smi05} find that the observed scaling relations for relaxed and not relaxed clusters are significantly different. However, \citet {oh06} reach the opposite conclusion, basing their analysis on simulated and observed data. \citet {lop06} find that the T$_X$-N$_{gals}$ (or T$_X$-L$_{opt}$) relation is severely affected by the presence of substructures, although the relations involving L$_X$ are not. 

The main goals of this paper are: (i) collect higher quality (SDSS) data for the NoSOCS supplemental catalog to redshift $z \sim$ 0.5 (ii) define a low-redshift subset of the NoSOCS sample ($z \le$ 0.1); (iii) derive unbiased measurements of velocity dispersion and masses for the clusters in this low$-z$ sample; (iv) compare optical and X-ray mass estimates, allowing checks for possible biases in the mass estimates. This paper gives a description of the estimation of cluster properties for the NoSOCS catalog using SDSS. New photometric redshifts, richness and optical luminosity are derived for the full sample. Velocity dispersion, mass, and physical radius are obtained for the low-$z$ clusters. This data set will be used for future cluster studies. 

This work relies on the supplement version of NoSOCS (\citealt {lop03, lop04}), which contains 9,956 cluster candidates over 2,700 square degrees. This catalog, based on the Digitized Second Palomar Observatory Sky Survey (DPOSS, \citealt {djo03}), includes clusters up to $z \sim 0.50$ and is the largest catalog to this redshift limit to date. Here, we discuss the use of higher quality photometric data from the SDSS (\citealt {yor00}) to assess NoSOCS cluster properties. We also used spectroscopic data from the SDSS to study the dynamics of the low redshift systems ($z \le 0.10$). 

The presentation of the paper is as follows: In $\S$2 we briefly describe the NoSOCS. Selection of SDSS data for the NoSOCS clusters is described in $\S$3, where we also discuss the estimates of richness and optical luminosity, as well as the effect of different k$-$corrections and evolution in M$^*$ to richness. In $\S$ 4 we detail the way we characterize the properties of the low-$z$ systems (velocity dispersion, characteristic radius, and mass). Specific issues regarding the method for interloper rejection, as well as the radial cutoff and magnitude limit of the spectroscopic survey are also discussed. We estimate X-ray luminosity for a subset of these local clusters and discuss the estimation of mass for a subset of systems with accurate T$_X$ available. In $\S,$ 5 we present the substructure tests employed for the clusters at low-$z$. We compare optical and X-ray mass estimates in $\S$ 6 (as well as optical masses from the caustic method and the virial analysis). We summarize our main conclusions in $\S$ 7. Throughout this work we assumed a cosmology with $\Omega_m = 0.3, \Omega_{\Lambda} = 0.7$ and H$_0 = 100$ $h$ $km$ $s^{-1}$ Mpc$^{-1}$, with $h$ set to 0.7.

\section{The Northern Sky Optical Cluster Survey (NoSOCS) AND CIRS}

The work presented here is based on the supplement version of the Northern Sky Optical Cluster Survey (NoSOCS, \citealt {lop03, lop04}). NoSOCS (\citealt {gal00, gal03, gal08}) is a catalog of galaxy clusters constructed from the digitized version of the Second Palomar Observatory Sky Survey (POSS-II; DPOSS, \citealt {djo03}). This catalog is derived from high-latitude fields $|b| > 30^\circ$, covering $\sim 11,000$ deg$^2$ and containing $\sim 15,500$ cluster candidates. Its construction is limited to $r = 19.5$, where the star/galaxy separation is reliable and the photometric errors are accurate enough to use the $g-r$ color as a redshift indicator. Object classification and photometric calibration are described in \citet {ode04} and \citet {gal04}, respectively. The supplement version of this survey (\citealt {lop04}) is a deeper cluster catalog ($z \leq 0.5$) within a smaller area of DPOSS ($\sim$ 2,700 square degrees), containing 9,956 cluster candidates. It covers a smaller area than NoSOCS because the search is restricted to the best DPOSS plates, selected according to seeing and limiting magnitude ($r = 21.0$). The selection is further restricted to high galactic latitude ($|b| > 50^\circ$), where stellar contamination is modest and nearly uniform. As this catalog is based on deeper galaxy catalogs, the larger photometric errors prevent the use of colors (from the plate material) for redshift estimates, which instead are simply based on a magnitude--redshift relation. 

In section 4 we derive cluster properties for low redshift ($z \le 0.1$) 
clusters. In this redshift range NoSOCS comprises only poor systems (with $\sigma < 700$ km/s). That results from the small volume probed by the NoSOCS supplement within the SDSS spectroscopic survey (at $z \le 0.1$). Hence, we decided to complement our sample with the one including massive systems studied by \citet {rin06}. Their sample comprises 74 X-ray selected clusters at $z \le 0.1$ within the SDSS DR4 (see their Table 1). This set is named Cluster Infall Regions in SDSS (CIRS). That is an optimal data set to extend our sample, as it contains more massive clusters than this work, being also based on SDSS data and limited to the same redshift range.

Seven clusters of their sample are common to ours; three are excluded for being too nearby ($z < 0.01$) and covering a very large angular area. SDSS data for the remaining sixty-four CIRS clusters were extracted in the same way as we did for the NoSOCS (section 3). The only difference is the use 
of the SDSS DR6, instead of DR5 (CIRS was incorporated into this work when DR6 was available). We refined the spectroscopic redshift estimates of these clusters and eliminated seven systems not well sampled within 2.5 h$^{-1}$ Mpc (inspection performed by PAAL). Another cluster (Zw1665) was  eliminated for having a large difference between the new and the CIRS redshift ($|z_{new} - z_{CIRS}| = 0.0163$) and because $z_{new} = 0.0139$ is too low to be included. The final CIRS sample used here contains fifty-six clusters.

\section{Selection of Sloan Digital Sky Survey (SDSS) data for the NoSOCS clusters}

The photometric and spectroscopic data used in this paper, were taken from the fifth release of the SDSS \citep {yor00}. We have selected only objects from the ``Galaxy'' view (so that only \texttt {PRIMARY} objects are allowed) to avoid duplicate observations. Standard flags for clean photometry are also used. As we select imaging and spectra, a {\it joined} query of the Galaxy and SpecObj views is performed. All the magnitudes retrieved from SDSS are de-reddened model magnitudes.

Before selecting photometric and spectroscopic SDSS data in the region of each NoSOCS cluster, we decided to clean the catalog in a more rigorous way than was done by \citet {lop04} where we eliminated ``double'' clusters by simply excluding objects within $60''$ of each other and located in the overlapping regions of different DPOSS plates. These ``double'' clusters are candidates with small separation (see \citealt {lop04}). The originally cleaned catalog contains 9,956 cluster candidates. A second cleaning of the NoSOCS catalog was done here, eliminating all clusters within 0.75 h$^{-1}$ Mpc (1.07 Mpc for $h = 0.7$) of each other and a photometric redshift difference of $\Delta z \le 0.05$ (the photometric redshift accuracy of the supplemental NoSOCS catalog). When two clusters match this criterion, we keep the richer cluster. From the 9,956 original systems 358 (3.6\%) objects are eliminated, leaving 9,598 clusters.

Considering the higher quality of the SDSS data compared to DPOSS, and the availability of many colors, we decided to obtain new photometric redshift estimates ($z_{photo}$) for all the NoSOCS systems found within SDSS. Thus, initially, for each NoSOCS cluster we selected SDSS data in circular areas with a radius of 3.0 Mpc + 0.5 degrees around the cluster center, for which we still consider the original photometric redshifts from the plate data. We found 7,822 of the NoSOCS clusters within the SDSS area. The limiting magnitude of the selected data is $r = 21.0$, the star/galaxy separation limit in SDSS. We have also retrieved data for fifty randomly selected blank fields of one square degree each. The data extraction for these fields is done in the same way as for the cluster areas. These control fields are used for background subtraction performed when estimating photometric redshifts, richness (N$_{gals}$) and optical luminosity (L$_{opt}$).

To obtain new photometric redshifts we adopted an empirical approach based on the measurement of the mean $r$ band magnitude and median (g-i) and (r-i) colors. The procedure for estimating $z_{photo}$ is described in section 4 of \citet {lop07}. 

With the more accurate photometric redshift estimates for each cluster, we repeated the selection of the galaxy data. This time we considered regions of 10.0 Mpc x 10.0 Mpc. Then, an inspection of finding charts for all cluster areas was performed by PAAL. We wanted to exclude clusters that were not well sampled, either because they were too close to the edges of the SDSS or had  a bright (masked) object near the center of the field. After this inspection we kept 7,550 clusters (96.5\%). Finally, we obtain a refined position for each cluster. This process of re-centering considers the luminosity-weighted coordinates of all galaxies within 0.30 h$^{-1}$ Mpc of each cluster (see \citealt {lop06}). This aperture is chosen to be large enough to improve the centroid estimate avoiding projection effects that would affect the new coordinates. Photometric redshift estimates from the new centroids are also determined for all objects, as the $z_{photo}$ may be sensitive to this parameter. Considering the more accurate redshifts, new N$_{gals}$ and L$_{opt}$ estimates are obtained for each cluster, as will be discussed in the next section. 

As $z_{photo}$ is greatly improved from the original value in NoSOCS we performed a final cleaning of the cluster catalog, to eliminate ``double'' clusters. The procedure is analogous to the one described in the beginning of this section. Based on the original coordinates and the associated $z_{photo}$, we eliminated clusters that are still within 0.75 h$^{-1}$ Mpc of each other and have a redshift offset of $\Delta z \le 0.025$ (taking in account the improved accuracy in $z_{photo}$). Again, we keep the richer cluster of two coincident ones. One hundred thirty-six (1.8\%) of the 7,550 are eliminated. The distribution of photometric redshifts for the 7,414 clusters is shown in Figure 1, while the richness distribution is exhibited in Figure 2. In agreement to the selection function of the catalog, we see that at high redshifts the sample is preferentially composed of rich clusters.

\begin{figure*}
\begin{center}
\leavevmode
\includegraphics[width=7.2in]{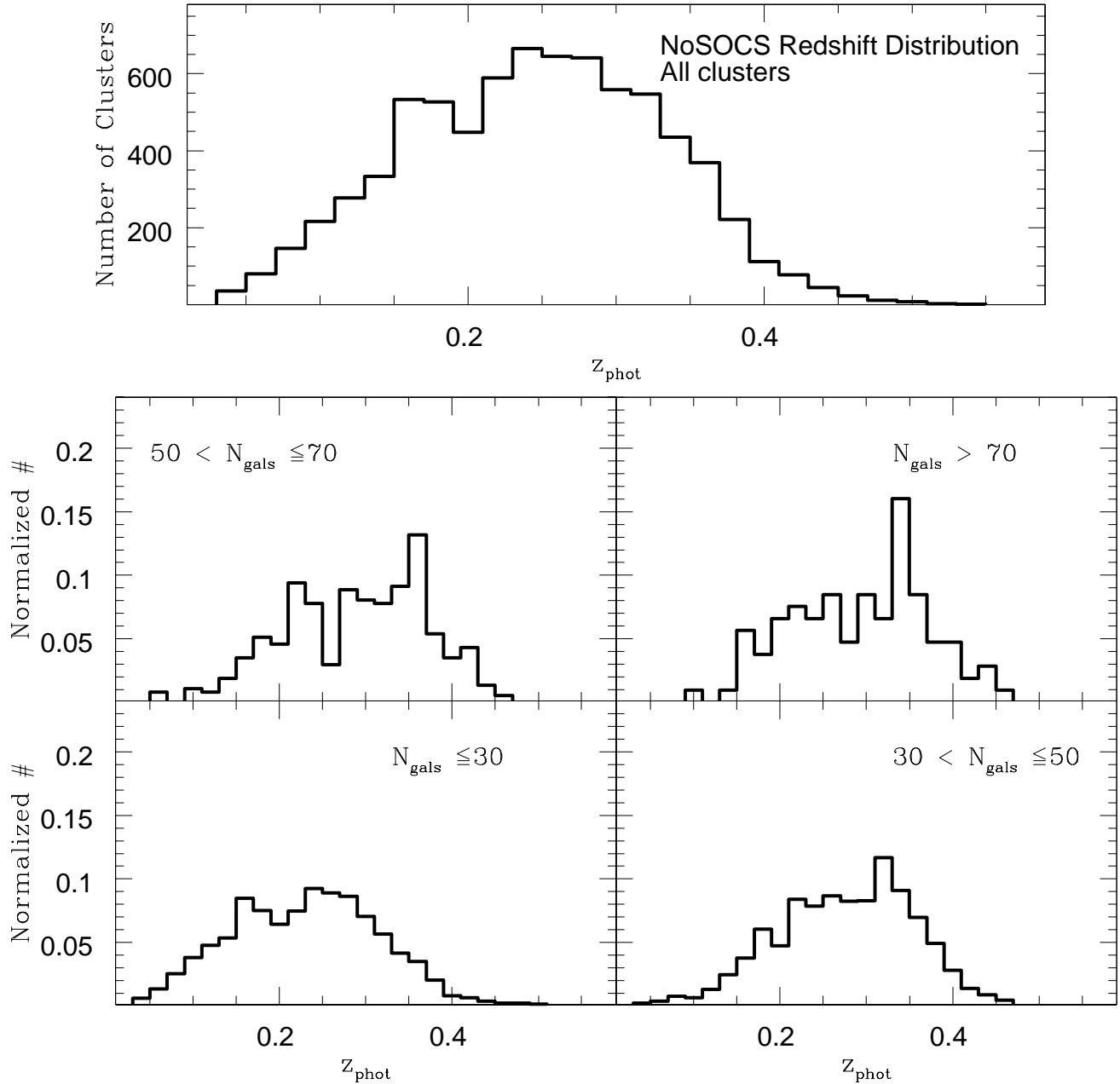}
\end{center}
\caption{Photometric redshift distributions of the 7,414 NoSOCS clusters for which we extract data from SDSS (top). The four bottom panels show the $z_{photo}$ distributions in different richness bins (indicated on 
each panel).}
\label{fig:zspeczphot}
\end{figure*}

\begin{figure*}
\begin{center}
\leavevmode
\includegraphics[width=7.2in]{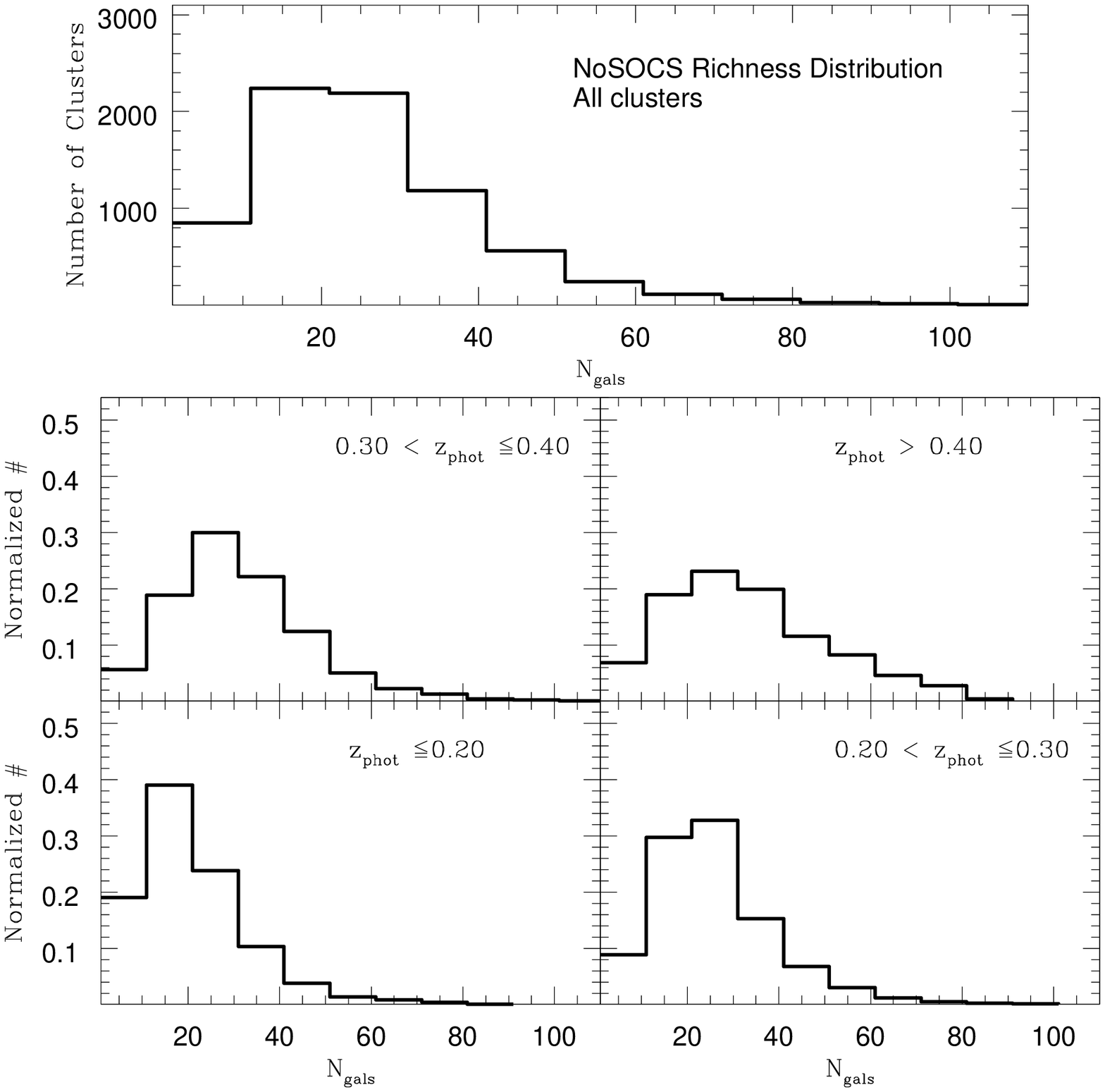}
\end{center}
\caption{Richness distributions of the 7,414 NoSOCS clusters for which we extract data from SDSS (top). The four bottom panels show the richness distributions in different $z_{photo}$ bins (indicated on each panel).}
\label{fig:zspeczphot}
\end{figure*}

\subsection{Richness and optical luminosity}

The procedure to estimate the richness of a cluster depends on the cluster centroid, radius adopted, and the redshift. A detailed description can be found in \citet {lop06}. Here, we highlight the main points. Our richness measure, $N_{gals}$, is the number of galaxies at $m^*_r - 1 \le m_r \le m^*_r + 2$ within a given aperture, where $m^*_r$ is the characteristic apparent magnitude of the cluster luminosity function. For this work we consider the bright end values of the Schechter cluster luminosity function (LF) obtained by \citet {pop06}, where their fits are optimized for bright and faint magnitudes simultaneously. For the bright end, they found $\alpha = -1.09$ and $M^* = -20.94$ within $R_{200}$. These values of M* and $\alpha$ are consistent with previous results. The value of M* is converted to the same cosmology used here and to the proper value at $z = 0$ (taking the mean redshift of their sample as $z = 0.1$). Then, for different redshifts we adopted a mild evolutionary correction to the value of M*, given by $M^*(z) = M^*(0) + Qz$, with Q = -1.4 \citep {yee99}.

As we do not have an estimate of a radius that scales with mass (such as $R_{200}$) we adopt the optimal aperture of 0.50 h$^{-1}$ Mpc (0.71 Mpc for $h = 0.7$). This choice results in the smallest errors in the richness estimates and in the most accurate scaling relations (between optical and X-ray properties). More details are found in \citet {lop06} (similar arguments are given by \citealt {pop04}, using the RASS-SDSS sample). k$-$corrections are applied to the magnitude of each galaxy, as explained in items (i) and (iii) below. These corrections are obtained through the convolution of spectral energy distributions from \citet {col80} with the SDSS $r$ filter. The steps to estimate richness are:

\begin{enumerate}
\item We use the photometric redshift of each cluster to convert an absolute magnitude $M^*_r$ to $m^*_r$ and to calculate the apparent radius (in seconds of arc) for a fixed aperture of 0.50 h$^{-1}$ Mpc. The photometric redshift is also used to compute the k$-$correction typical of elliptical and late-type galaxies (Sbc). These are named ``ke'' and ``ks'', respectively. Since we count only galaxies with $m^*_r - 1 \le m_r \le m^*_r + 2$, initially we select all of them within 0.50 h$^{-1}$ Mpc of the cluster center at $m^*_r - 1 + ks$ $ \le m_r \le$ $m^*_r + 2 + ke$.  The k$-$corrections are applied to individual galaxies at a later stage, so these limits guarantee that we select all galaxies that can fall within $m^*_r - 1 \le m_r \le m^*_r + 2$. N$_{clu}$ is the number of galaxies selected in the cluster region.

\item We use the fifty blank fields of 1 square degree to estimate the background counts to be subtracted from the cluster counts. Galaxies are selected within the same magnitude range as used for computing N$_{clu}$, and the counts are scaled to the same area of the cluster. The median counts from all fifty boxes give the background estimate, N$_{bkg}$; (we actually exclude boxes with counts outside the range $\pm 3 \sigma$ of the median counts). We adopt the interquartile range (IQR, which is the range between the first and third quartiles) as a measure of the error in N$_{bkg}$, which we term ${\sigma_{bkg}}$ (for normally distributed data IQR = 1.35 $\sigma$, where $\sigma$ is the standard deviation). The background corrected cluster counts (N$_{clu}$ - N$_{bkg}$) is called N$_{corr}$.

\item Next, we apply k$-$corrections to the galaxy populations in each cluster using a bootstrap procedure. In each of the one-hundred realizations, we randomly select N$_{corr}$ galaxies from those in the cluster region (N$_{clu}$). We then apply a k$-$correction to the magnitude of each galaxy. An elliptical k$-$correction is applied to X$\%$ of the N$_{corr}$ galaxies, while a Sbc k$-$correction is applied to the remaining (100-X)$\%$. Then, we use these k$-$corrected magnitudes to count the number of galaxies at $m^*_r - 1 \le m_r \le m^*_r + 2$. The final richness estimate N$_{gals}$ is given by the median counts from the one-hundred realizations. We considered a variable fraction of galaxy type at different redshift ranges. At $z \le 0.15$ we assumed clusters are composed of 80\% ellipticals; 50\% at $0.15 < z \le 0.30$ and 30\% at $z > 0.30$. Different values for these fractions have little effect on the final richness estimates. The richness error from the bootstrap procedure alone is given by ${\sigma_{boot}}$ = IQR and the total error is the combination of this error and the background contribution, so that ${\sigma_{N_{gals}}}$ = $\sqrt{{\sigma_{boot}}^2 + {\sigma_{bkg}}^2}$. From now on, we simply call the richness error $N_{gals-err}$.

\item If the cluster is too nearby or too distant, either the bright ($m^*_r - 1 + ks$) or the faint ($m^*_r + 2 + ke$) magnitude limit will exceed one of the survey limits ($10.0 \le m_r \le 21.0$). Here, we apply one of the following correction factors to the richness estimate:

\bea
{\rm \gamma_1} = {\int_{m_r^*-1}^{m_r^*+2} \Phi(m)dm \over
\int_{10}^{m_r^*+2} \Phi(m)dm} \,\,\,
\label{eq:perpdef}
\eea

\bea
{\rm \gamma_2} = {\int_{m_r^*-1}^{m_r^*+2} \Phi(m)dm \over
\int_{m_r^*-1}^{21} \Phi(m)dm} \,\,\,
\label{eq:perpdef}
\eea
We call $\gamma_1$ and $\gamma_2$ the low and high magnitude limit correction factors. Whenever necessary, one of the above factors is multiplied by N$_{gals}$ and $N_{gals-err}$.

\end{enumerate}

The photometric redshift and richness distributions are illustrated in Figures 1 and 2. Note that the distributions use richness estimates without evolution in the value of M$^*$. Such evolution would make the richness estimate smaller and would cause the distributions to be shifted to the left. These differences affect only clusters at higher redshifts ($z >$ 0.25). Although the listed values in the tables consider evolution in M$^*$, the mass estimates and scaling relations, obtained only at $z \le$ 0.10, are not affected by this correction. 

The DR5 of SDSS also provides photometric redshifts, galaxy types and k$-$corrections for all galaxies obtained through a template fitting procedure. Thus, we can test if the richness estimates are sensitive to the k$-$correction used, if individual or based on a statistical approach. We consider the individual k$-$correction and \texttt {zphot} of each galaxy to obtain new estimates of richness. The procedure is analogous to that used above, except for the k$-$correction that is applied to each galaxy (see item iii above). In all the one-hundred realizations, when selecting N$_{corr}$ galaxies we apply the k$-$correction given by SDSS to each galaxy, avoiding the need to statistically consider different galaxy types. However, as we can see from Figure 3, the richness estimates obtained in both cases are indeed very similar.

To estimate the optical luminosity of each cluster, we proceed as follows. We generate the magnitude distribution (assuming a bin of $\Delta mag = 0.2$) of the N$_{corr}$ galaxies found within the cluster regions, at each magnitude bin. Then, the corrected optical luminosity is given by

\bea
{\rm L} = \sum_1^n N_i 10^{-0.4r_i} \,\,\,,
\label{eq:perpdef}
\eea
where N$_i$ represents the corrected counts at each magnitude bin r$_i$. The sum is performed over all $n$ magnitude bins. This luminosity is converted to an apparent magnitude associated to the cluster, from which we subtract an elliptical k$-$correction, $m_{clu} = -2.5log (L) - ke$. We then consider the distance modulus (DM) for the cluster redshift and transform this value to an absolute magnitude, $M_{clu} = m_{clu} - DM$. Finally, considering the absolute magnitude of the Sun in the $r$ band of SDSS ($M_{\odot}^r = 4.64$;  \citealt {bla07}), we obtain the optical luminosity in solar units as L$_{opt} = 10^{-0.4(M_{clu} - M_{\odot}^r)}$. The error in the luminosity is given by 

\bea
\Delta L = \sqrt{{\sum_1^n [(N_i + \sigma_{ibkg}^2)^{1/2}10^{-0.4r_i}]^2}} \,\,\, ,
\label{eq:perpdef}
\eea
where $\sqrt {N_i}$ and $\sigma_{ibkg}$ are the errors in the cluster corrected counts and background in the {\it i-th} magnitude bin. The term in brackets represents the combined error in each bin. If the survey limits are exceeded, one of the correction factors below is multiplied by L (equation 3) and $\Delta L$ (equation 4).

\bea
{\rm \Gamma_1} = {\int_{m_r^*-1}^{m_r^*+2} \Phi(m)mdm \over
\int_{10}^{m_r^*+2} \Phi(m)mdm} \,\,\,
\label{eq:perpdef}
\eea

\bea
{\rm \Gamma_2} = {\int_{m_r^*-1}^{m_r^*+2} \Phi(m)mdm \over
\int_{m_r^*-1}^{21} \Phi(m)mdm} \,\,\,
\label{eq:perpdef}
\eea
\begin{figure}
\includegraphics[height=0.50\textwidth,angle=0]{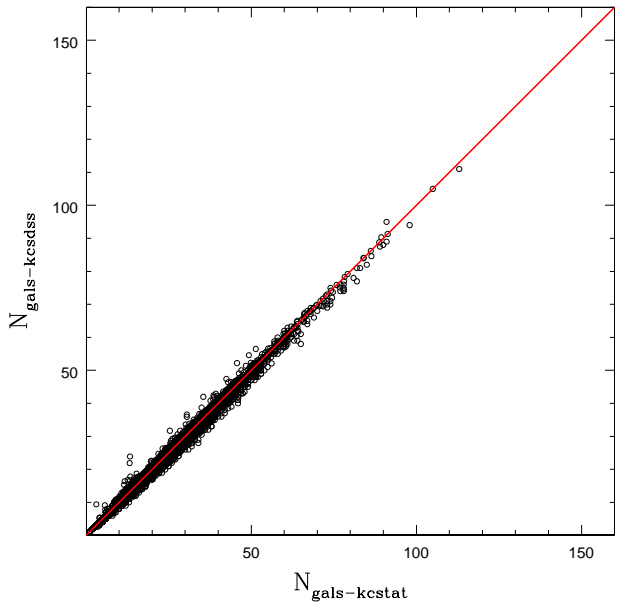}
\caption{Comparison between richness estimates obtained with a statistical k$-$correction (horizontal axis) and using the values obtained for each galaxy in SDSS (vertical axis). The solid line (red in the electronic edition) indicates the Y = X result.}

\label{fig:richcomp}
\end{figure}

\section{Properties of low redshift clusters}

In this section, we describe the procedures adopted for estimating cluster properties such as velocity dispersion ($\sigma$), characteristic radius ($R_{500}$ and $R_{200}$) and mass ($M_{500}$ and $M_{200}$). These parameters are measured only for clusters at $z \le 0.10$. We use the SDSS spectroscopic survey to derive these estimates. This survey is complete to $z \sim 0.10$, where the peak of the redshift distribution is located. At higher redshifts we start missing galaxies fainter than $M^*+1$, and that may bias the dynamical analysis. Details regarding this choice are given below. It is also important to stress that some of the more recent cluster catalogs from SDSS are not always well sampled at $z \le 0.10$. For instance, the MaxBCG catalog was constructed at $0.10 \le z \le 0.30$. 

The first step to determine velocity dispersion, characteristic radius, and mass is the identification of the cluster spectroscopic redshift and velocity limits, as well as the selection of cluster members from the spectroscopic sample. We start by identifying groups in redshift space and thus obtain a spectroscopic redshift for each cluster. We use the gap-technique described in \citet {kat96} and \citet {ols05} to identify gaps in the redshift distribution. Instead of adopting a fixed gap, such as 1000 $km$ $s^{-1}$ or $0.005(1+z)$, we considered a variable gap, called {\it density gap} \citep {ada98}. The density gap size is given by the expression $\Delta z = 500(1+\exp(-(N-6)/33))/c$, where N is the number of galaxies found in the redshift survey of a cluster \citep {ada98}, and $c$ is the speed of light in $km/s$. Here, we select all galaxies within 0.50 h$^{-1}$ Mpc of the cluster center. When performing galaxy selection, the only requirement is that the galaxy is brighter than $r =18.0$ (approximately the magnitude limit of the SDSS spectroscopic survey). In the region of each cluster, groups in $z$-space are identified through the application of the gap-technique. The redshift of each group is given by the biweight estimate \citep {bee90}.

The next step is to assess the significance of each of these groups. For that we consider the full spectroscopic survey of SDSS (considering the same limit at $r =18.0$). For a given cluster, for each group identified within it, we draw 1000 sets of galaxies. These sets have the same number of galaxies as in the cluster region where the group was identified. The gap-technique is applied exactly as before, and then we check the probability of finding groups with at least the same number of galaxies at the redshift of the original group. We consider only groups that are significant at the 95\% level. 

From all the significant groups (sometimes there is only one), we select the one that has the smallest redshift difference to the photometric value of the cluster. We also require this difference ($|z_{photo} - z_{spec}|$) to be smaller than $0.03(1+z_{spec})$, where $z_{photo}$ is the photometric redshift estimate of the cluster ($\S$ 3.1) and $z_{spec}$ the spectroscopic value of the group. Finally we also require that the group have at least three member galaxies within the 0.50 h$^{-1}$ Mpc aperture (most have many more). During this procedure, we also obtain the velocity (or redshift) limits ($v_{lo}$ and $v_{hi}$) of the cluster within this radius. This procedure is applied to all 754 clusters (out of the original 7,414) with $z_{photo} \le$ 0.133. From those we find 179 clusters with $z_{spec} \le 0.10$ (having at least 3 galaxies within 0.50 h$^{-1}$ Mpc). From now on, we will refer to $z_{spec}$ as $z_{cluster}$. We noticed that for two systems the velocity dispersion estimates (see below) were artificially high because of the influence of a projected cluster along the line-of-sight. We found that a refined redshift estimate (within 0.30 h$^{-1}$ Mpc, instead of 0.50 h$^{-1}$ Mpc) helps solve this problem.

Next we apply an algorithm for interloper rejection to arrive at a final list of cluster members, which will be used to estimate the cluster properties. The procedure we adopt is similar to that of \citet {fad96}  and is called the ``shifting gapper'' technique. It works through the application of the gap-technique in radial bins from the cluster center. The bin size is 0.42 h$^{-1}$ Mpc (0.60 Mpc for h = 0.7) or larger to force the selection of at least 15 galaxies (consistent with \citealt {fad96}). Galaxies not associated with the main body of the cluster are eliminated. This procedure is repeated until the number of cluster members is stable (no more galaxies are rejected as interlopers). This method makes no hypotheses about the dynamical status of the cluster, while other procedures are based on physical assumptions about the cluster mass profile (see \citealt {woj07}). The way we apply this technique has many details, which are described below.

\begin{enumerate}
\item We select all galaxies within 2.5 h$^{-1}$ Mpc (3.57 Mpc for $h = 0.7$) from the cluster center and with $|cz - cz_{cluster}| \le 4000$ $km$ $s^{-1}$. The aperture of 2.5 h$^{-1}$ Mpc is typically larger than the virial radius of clusters, but is still small enough to avoid the influence of large scale structure on the cluster analysis. More details about this choice are given below.

\item Before starting the shifting gapper procedure, we take the maximum velocity offset and impose the velocity limits to be symmetric. The velocity offsets within 0.50 h$^{-1}$ Mpc are $|v_{lo}-v_{cluster}|$ and $|v_{hi}-v_{cluster}|$, where $v_{cluster} = cz_{cluster}$. The maximum offset between these two is named $v_{diff}$ and the new symmetric velocity limits are defined as $v_{lo} = v_{cluster} - v_{diff}$ and $v_{hi} = v_ {cluster} + v_{diff}$. If $v_{diff} < 300(1+z_{cluster})$ $km$ $s^{-1}$ we set it to this value. We have also noticed that the factor ``500'' in the density gap expression ($\Delta z = 500(1+\exp(-(N-6)/33))/c$) may not be suitable when running the shifting gapper for low and high mass systems in radial bins. Thus for each radial bin we define the parameter $facgap = |v_{lo}-v_{hi}|/10$ and modify the expression to be $\Delta z = facgap(1+\exp(-(N-6)/33))/c$. This parameter $facgap$ assumes a minimum value of 300. 

\item When we select galaxies within a radial bin of 0.42 h$^{-1}$ Mpc or larger (in the case less than fifteen galaxies are found), if the radial offset between two consecutive galaxies is greater than 0.7 h$^{-1}$ Mpc, the procedure is halted. Then all galaxies with a radial distance from the cluster center greater or equal to this galaxy's radial distance are eliminated as interlopers.

\item When starting the procedure, in the first radial bin, we define $v_{lo}$, $v_{hi}$ and $facgap$ as above. Then we run the gap technique for the galaxies of that radial bin. If at least one group is identified in redshift space, we sort these in velocity offset from the cluster. We consider the first group (smallest velocity offset) valid if the offset is $\le c(0.001(1+z_{cluster}))$. We then consider all the galaxies from this group as cluster members, except those outside the velocity limits, $v_{lo}$ and $v_{hi}$. If the first group offset is larger than $c(0.001(1+z_{cluster}))$, or if there are no groups found in this radial bin, we still keep all galaxies within $v_{lo}$ and $v_{hi}$. If the first group is considered as valid, we also take the velocity limits of this group to update the values of $v_{lo}$ and $v_{hi}$, exactly as above, keeping these values symmetric and varying with radial distance. For the next radial bin, these new values of $v_{lo}$ and $v_{hi}$ will be considered and so on. 

\end {enumerate}

After the interloper removal we kept one-hundred and twenty-seven clusters (out of one-hundred and seventy-nine) with at least ten member galaxies selected. Figure 4 illustrates the result of the cluster member selection for these 127 NoSOCS systems. On each panel one galaxy cluster is represented, with filled squares showing cluster members and open circles galaxies rejected as interlopers.

\begin{figure*}
\begin{center}
\leavevmode
\includegraphics[width=7.2in]{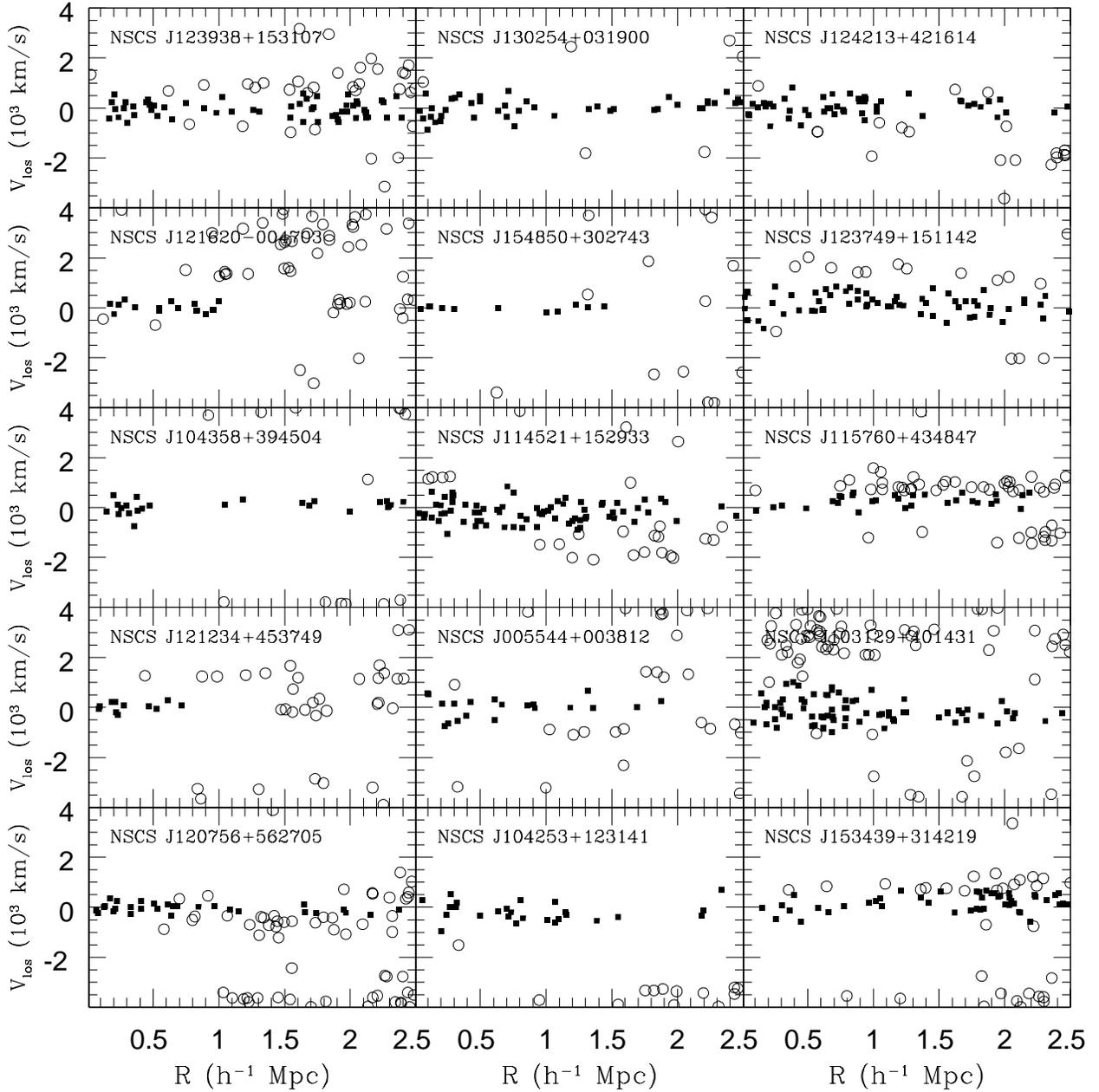}
\end{center}
\caption{Phase-space diagrams of 15 NoSOCS clusters shown as examples. The velocity and radial offsets are with respect to the cluster center. We apply a shifting gapper procedure for the selection of cluster members (filled squares) and exclusion of interlopers (open circles). Similar diagrams for all NoSOCS clusters are available in the electronic edition of the MNRAS.} 
\label{fig:zspeczphot}
\end{figure*}

We then perform a virial analysis of all the one-hundred and twenty-seven clusters with at least ten galaxies (trying to use systems with less than ten galaxies leads to increased scatter in the scaling relations). The procedure is analogous to what is done by \citet {gir98}, \citet {pop05, pop07} and \citet {biv06}. First, we compute the line-of-sight velocity dispersion ($\sigma_{P}$) of all cluster members within the aperture R$_A$, which is the radial offset of the most distant cluster member. R$_A$ is usually close to 2.5 h$^{-1}$ Mpc (the maximum radius adopted for this work), but could be much smaller if the outermost galaxies are considered interlopers. The robust velocity dispersion estimate ($\sigma_{P}$) is given by the gapper or biweight estimator, depending on whether the number of objects available is $< 15$ (gapper) or $\ge 15$ (biweight; \citealt {bee90}). Following the prescriptions of \citet {dan80}, the velocity dispersion is corrected for velocity errors. We also obtain an estimate of the projected ``virial radius'' R$_{PV}$ \citep {gir98}. A first estimate of the virial mass is given by (equation 5 of \citealt {gir98})

\bea
{\rm M_V} = {3\pi \sigma^2_{P} R_{PV} \over
             2 G},\,\,\,
\label{eq:perpdef}
\eea
where G is the gravitational constant and $3\pi/2$ is the de-projection factor. 

Next we apply the surface pressure term correction to the mass estimate \citep {the86}. To do that we first estimate the Navarro et al. (1997, NFW from now on) concentration parameter, using the relation $c = 4 \times (M/M_{KBM})^{-0.102}$ (see also \citealt {pop07, biv06}). The slope is taken from \citet {dol04} and the normalization ($M_{KBM} \approx 2 \times 10^{15} M_{\odot}$) is from \citet {kat04}. To apply the surface pressure term we also need an estimate of the R$_{200}$ radius, which (as a first guess) is taken from the definition of Carlberg et al. (1997; equation 8 of that paper). The correction (computed as in equation 14 of \citealt {gir98}) is applied considering isotropic orbits. In particular in our analysis, the term associated to the presence of velocity anisotropies is set to 0.3, as found by \citet {gir98} at R$_{vir}$. We noticed that for different velocity dispersion profiles, and at other radii, different values should be adopted for this term. That may lead to differences of up to $\sim 25\%$ in the value of the surface pressure correction, but the effect on the mass is less than 5\%. Since the effect is much smaller than the typical error in mass, and the study of velocity anisotropies is beyond the scope of this paper; we consider the value of 0.3 suitable. Note also that the most important issue here is to apply the surface pressure correction, which is sometimes neglected in the literature. 

After applying this correction, we obtain a refined estimate of R$_{200}$ considering the virial mass density. If $M_V$ is the virial mass (after the surface pressure correction) in a volume of radius $R_{A}$, R$_{200}$ is then defined as R$_{200} = R_{A}[\rho_{V}/(200\rho_c(z))]^{1/2.4}$. In this expression $\rho_{V} = 3M_V/(4\pi R_{A}^3)$ and $\rho_c(z)$ is the critical density at redshift $z$. The exponent in this equation is the one describing an NFW profile near R$_{200}$ \citep {kat04}. If we use five-hundred instead of two-hundred on this equation, we obtain an estimate of R$_{500}$. Next, assuming an NFW profile we obtain M$_{200}$ (or M$_{500}$) from the interpolation (most cases) or extrapolation of the virial mass M$_V$ from R$_{A}$ to R$_{200}$ (or R$_{500}$). Then, we use the definition of M$_{200}$ (or M$_{500}$) and a final estimate of R$_{200}$ (or R$_{500}$) is derived. This procedure is analogous to what is done by \citet {pop07} and \citet {biv06}. We assume the percentage errors for the mass estimates are the same before and after the surface pressure correction (as in \citealt {gir98}).

Table 1 lists the one-hundred and twenty-seven NoSOCS clusters and their main characteristics. The cluster name is in column 1; coordinates are shown in columns 2 and 3; redshift in 4; richness and its error in column 5; optical luminosity and its error in column 6; and the value of the $\beta$ and $\Delta$ parameters obtained from the substructure tests and their significance (see section 5) are given in columns 7-10. Table 2 lists the velocity dispersion, number of cluster members used for that estimate, number of cluster members within R$_{200}$, and the  characteristic radii and masses of these clusters, R$_{500}$, M$_{500}$, R$_{200}$, and M$_{200}$.

Table 3 lists the main characteristics of the CIRS systems. The same parameters, as given in Table 1, are shown here plus the original redshift listed in the CIRS (fourth column). Table 4 is analogous to Table 2, but for the fifty-six CIRS clusters. Phase-space diagrams for these fifty-six clusters are shown in Figure 5. Note that the one-hundred and twenty-seven NoSOCS clusters have velocity dispersion estimates of $100 < \sigma < 700$ km/s, while the CIRS systems have $200 < \sigma < 900$ km/s (with only 23\% of objects with $\sigma < 400$ km/s). We noticed that three CIRS systems still have a large difference between the original and new redshifts, but not as large as Zw1665. For one of these three clusters, we find that a new redshift estimate within 0.3 h$^{-1}$ Mpc minimizes the influence of a neighbor system (as done for two NoSOCS clusters; see above). These three clusters are outliers in the scaling relations (paper II). However, they are still listed in Tables 3 and 4. These three clusters are Abell 1035B, Abell 1291A, and Abell 1291B (see Figure 5). 

\begin{table*}
\begin{minipage}{165mm}
\caption{Main properties of the 127 NoSOCS clusters at $z \le$ 0.01 and with at least 10 member galaxies.}
\label{tab:nosocs127}
\begin{tabular}{@{}lcccccccccc}
\hline
 name & ra & dec & z$_{\odot}$ & N$_{gals}$ & L$_{opt}$ & $\beta$ & $\beta_{sig}$ & $\Delta$ & $\Delta_{sig}$ \\
 & (J2000) & (J2000) & & & (10$^{12}$L$_{\odot}$) & & & & \\

\hline
NSCS J121847+484410 & 184.67720 & 48.72197 & 0.0448 & 10.0 $\pm$  3.0 & 0.229 $\pm$  0.093 & 109.600 & 0.048 & 6.577 & 0.717 \\
NSCS J011502+002441 & 18.75351 & 0.37955 & 0.0449 & 66.0 $\pm$  14.5 & 1.256 $\pm$  0.213 & 46.200 & 0.000 & 79.405 & 0.317 \\
NSCS J100242+324218 & 150.67372 & 32.70033 & 0.0505 & 29.0 $\pm$  7.8 & 0.351 $\pm$  0.108 & -14.400 & 0.460 & 44.065 & 0.571 \\

\hline
\end{tabular}
{Note. - A portion of this table is shown here for guidance regarding 
its form and content. A full version is available in the electronic
edition of the MNRAS.}
\end{minipage}
\end{table*}

\begin{table*}
\begin{minipage}{155mm}
\caption{Velocity dispersion, characteristic radii and masses of 127 
NoSOCS clusters.}
\label{tab:nosocs127}
\begin{tabular}{@{}lccccccc}
\hline
 name & $\sigma_P$ & N$_{obj}$-$\sigma_P$ & N$_{obj}$-R$_{200}$ & R$_{500}$ & M$_{500}$ & R$_{200}$ & M$_{200}$ \\
 & (km s$^{-1}$) & & & (Mpc) & (10$^{14}$M$_{\odot}$) & (Mpc) & (10$^{14}$M$_{\odot}$) \\
\hline
NSCS J121847+484410 & 179.9 $_{-34.7} ^{+59.1}$ & 29 & 12 & 0.49 $_{-0.06} ^{+0.11}$ & 0.35 $_{-0.14} ^{+0.23}$ & 0.67 $_{-0.09} ^{+0.15}$ & 0.36 $_{-0.14} ^{+0.24}$ \\
NSCS J011502+002441 & 474.6 $_{-28.0} ^{+36.2}$ & 118 & 78 & 1.05 $_{-0.04} ^{+0.05}$ & 3.46 $_{-0.41} ^{+0.53}$ & 1.45 $_{-0.06} ^{+0.07}$ & 3.63 $_{-0.43} ^{+0.56}$ \\
NSCS J100242+324218 & 376.1 $_{-31.7} ^{+40.8}$ & 78 & 46 & 0.78 $_{-0.04} ^{+0.06}$ & 1.41 $_{-0.24} ^{+0.31}$ & 1.07 $_{-0.06} ^{+0.08}$ & 1.45 $_{-0.25} ^{+0.32}$ \\

\hline
\end{tabular}
{Note. - A portion of this table is shown here for guidance regarding
its form and content. A full version is available in the electronic
edition of the MNRAS.}
\end{minipage}
\end{table*}

\begin{table*}
\begin{minipage}{165mm}
\caption{Main properties of the 56 CIRS clusters at $z \le$ 0.01 and with at least 10 member galaxies.}
\label{tab:cirs56}
\begin{tabular}{@{}lccccccccccc}
\hline
 name & ra & dec & z$_{\odot}$ & z$_{\odot}$ & N$_{gals}$ & L$_{opt}$ & $\beta$ & $\beta_{sig}$ & $\Delta$ & $\Delta_{sig}$ \\
 & (J2000) & (J2000) & (CIRS) & (New-SDSS) & & (10$^{12}$L$_{\odot}$) & & & & \\
\hline
MKW4 & 181.10500 & 1.90060 & 0.0204 & 0.0201 & 27.0 $\pm$  8.1 & 0.461 $\pm$  0.168 & -8.400 & 0.540 & 106.903 & 0.052 \\
MKW11 & 202.38361 & 11.78920 & 0.0228 & 0.0229 & 37.0 $\pm$  8.2 & 0.519 $\pm$  0.121 & -4.700 & 0.402 & 62.810 & 0.140 \\
A0779 & 139.92200 & 33.76300 & 0.0233 & 0.0231 & 33.0 $\pm$  7.9 & 0.546 $\pm$  0.120 & 6.000 & 0.194 & 107.662 & 0.138 \\

\hline
\end{tabular}
{Note. - A portion of this table is shown here for guidance regarding its form and content. A full version is available in the electronic edition of the MNRAS.}
\end{minipage}
\end{table*}

\begin{table*}
\begin{minipage}{135mm}
\caption{Velocity dispersion, characteristic radii and masses of the 56 
CIRS clusters.}
\label{tab:cirs56}
\begin{tabular}{@{}lccccccc}
\hline
 name & $\sigma_P$ & N$_{obj}$-$\sigma_P$ & N$_{obj}$-R$_{200}$ & R$_{500}$ & M$_{500}$ & R$_{200}$ & M$_{200}$ \\
 & (km s$^{-1}$) & & & (Mpc) & (10$^{14}$M$_{\odot}$) & (Mpc) & (10$^{14}$M$_{\odot}$) \\
\hline
MKW4 & 392.4 $_{-29.0} ^{+42.8}$ & 135 & 81 & 0.88 $_{-0.04} ^{+0.06}$ & 1.97 $_{-0.29} ^{+0.43}$ & 1.21 $_{-0.06} ^{+0.09}$ & 2.04 $_{-0.30} ^{+0.45}$ \\
MKW11 & 359.4 $_{-18.2} ^{+26.3}$ & 119 & 55 & 0.91 $_{-0.03} ^{+0.04}$ & 2.19 $_{-0.22} ^{+0.32}$ & 1.25 $_{-0.04} ^{+0.06}$ & 2.27 $_{-0.23} ^{+0.33}$ \\
A0779 & 400.2 $_{-24.8} ^{+37.4}$ & 135 & 81 & 0.89 $_{-0.04} ^{+0.06}$ & 2.07 $_{-0.26} ^{+0.39}$ & 1.23 $_{-0.05} ^{+0.08}$ & 2.15 $_{-0.27} ^{+0.40}$ \\

\hline
\end{tabular}
{Note. - A portion of this table is shown here for guidance regarding its form and content. A full version is available in the electronic edition of the MNRAS.}
\end{minipage}
\end{table*}

\begin{figure*}
\begin{center}
\leavevmode
\includegraphics[width=7.2in]{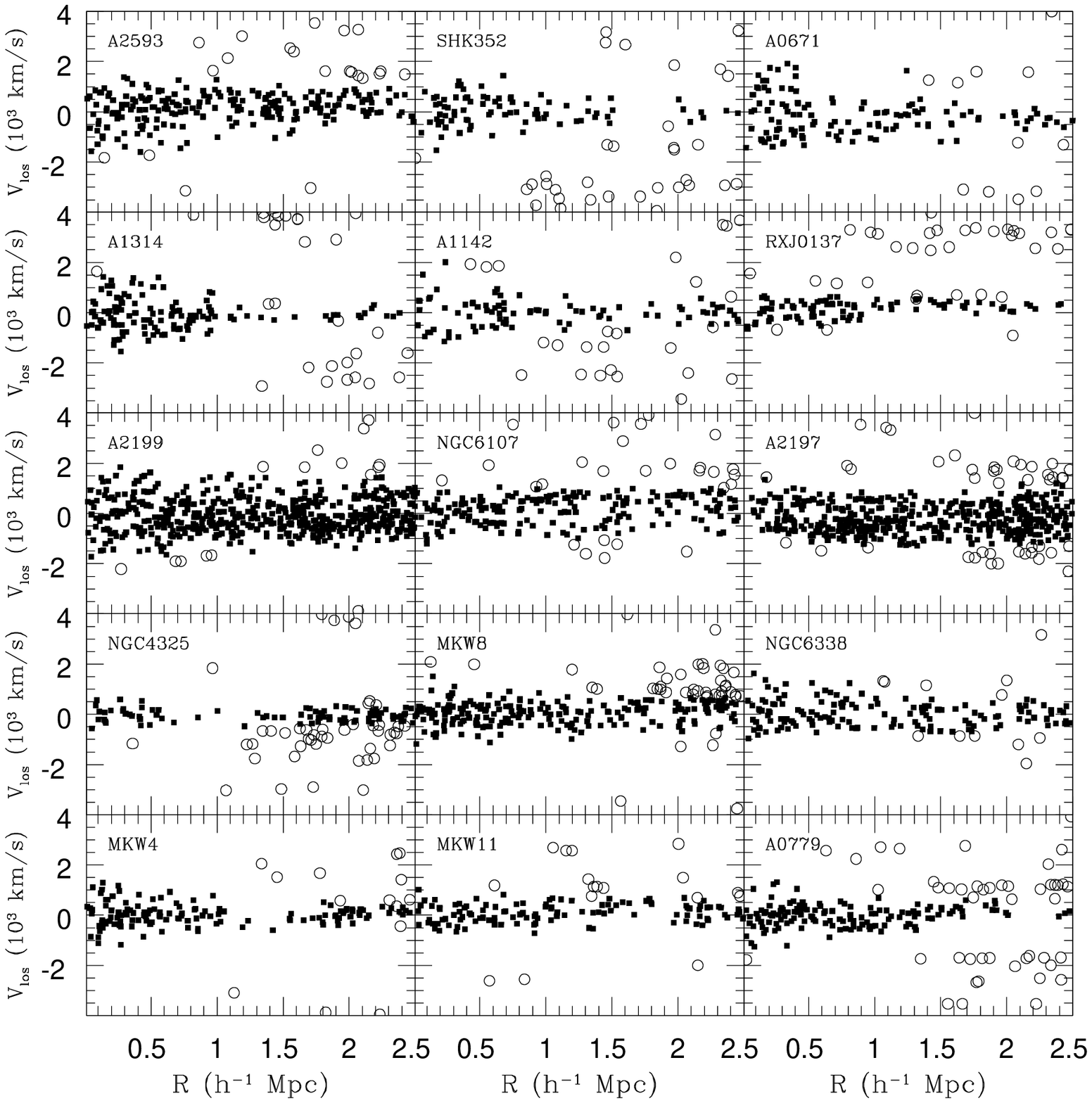}
\end{center}
\caption{Phase-space diagrams of fifteen CIRS clusters are shown here. The velocity and radial offsets are respective to the cluster center. We apply a shifting gapper procedure for the selection of cluster members (filled  squares) and exclusion of interlopers (open circles). Similar diagrams for all CIRS clusters are available in the electronic edition of the MNRAS.} 
\label{fig:zspeczphot}
\end{figure*}

In the next three subsections, we discuss a few issues that could affect the estimates of cluster properties. We focus on the interloper rejection and radius and luminosity limits of the spectroscopic survey. 

\subsection{Testing the rejection of interlopers}

In the literature, we find several different approaches for interloper removal in the dynamical modeling of galaxy clusters. For a recent comparison of the performance of several different methods applied to $N-$body cosmological simulations see the work of \citet {woj07}. As noted by these authors, differences in mass estimates may be explained by the number of interlopers a given method selects. That could also be an explanation for discrepant estimates to other methods based on X-ray observations or lensing analysis. Here, we do not intend to perform a detailed comparison. We use a method (shifting gapper) that has the following analysis advantages: use of the combined information of position and velocity; and is independent of any hypotheses regarding the dynamical status of the cluster. However, it is important to mention that even slight modifications in the procedure described above could lead to different estimates of the cluster properties. For instance, had we decided not to update the velocity limits $v_{lo}$ and $v_{hi}$ from one radial bin to the next, we would end up with fewer interlopers (see item iv of previous section). Then we would keep the values of $v_{lo}$ and $v_{hi}$ found within 0.50 h$^{-1}$ Mpc, instead of making those smaller for larger radius. If we had also enlarged these initial limits by a 3\% factor (to account for uncertainties in the choice of these limits), the number of interlopers would also decrease. For some clusters, the velocity dispersion could be $\sim 20\%$ larger and the mass $\sim 50\%$. For comparison, Figure 6 shows the phase-space diagrams for one cluster as an example. In the top panel, we show the same distribution as in Figure 4 and in the bottom we consider the velocity limits (within 0.50 h$^{-1}$ Mpc) to be larger by a factor of 3\% and do not update those for larger radial bins. As we can see, the number of interlopers (open circles) is now smaller. We decided to keep the approach described in the previous section because it is more stable at larger radii (especially for low mass systems). The scaling relations (see \citealt {lop08}; paper II) obtained with this approach are also found to have smaller scatter.

\begin{figure}
\begin{center}
\leavevmode
\includegraphics[width=3.5in]{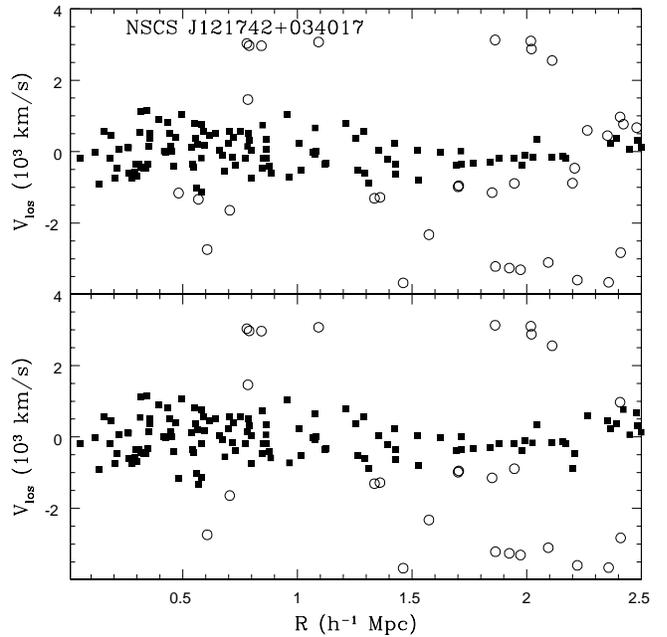}
\end{center}
\caption{Phase-space space diagrams of the cluster NSCS J121742+034017. In the 
top panel we show the original results (same as in Figure 4), while in the 
bottom we show the distribution when considering a less rigorous criteria for
interloper removal (see section 4.1). Although the differences in the final
number of galaxies rejected (open circles) are small, the larger number of 
cluster members (filled squares) results in larger estimates of velocity 
dispersion and mass. In particular, in a few cases those estimates could 
differ by more than 50\%.}
\label{fig:zspeczphot}
\end{figure}

\subsection{Radial cut-off}

One can consider as physically meaningful only the values of velocity dispersion found after considering all cluster galaxies within a sufficiently large radius. This issue has been previously discussed by \citet {fad96}, \citet{gir96} and \citet{gir98}. Here, we investigate how the estimates of $\sigma_{P}$ are affected by the maximum aperture (R$_{max}$) considered. In Figure 7, we show the comparison of $\sigma_{P}$ obtained with R$_{max} = 2.5$ h$^{-1}$ Mpc (horizontal axis) and R$_{max} =$ 0.5, 1.5 and 3.5 h$^{-1}$ Mpc (vertical axis). This figure clearly shows that a small aperture (0.5 h$^{-1}$ Mpc) tends to overestimate the velocity dispersion. At larger apertures, the effect is more subtle. We use R$_{max} = 2.5$ h$^{-1}$ Mpc because it is typically larger than the virial radius of clusters and still small enough to avoid the influence of large scale structure or nearby clusters. From inspection of the phase-space diagrams of several clusters, we found this aperture to be a good choice when working both with low and high mass systems.

\begin{figure}
\includegraphics[height=0.50\textwidth,angle=0]{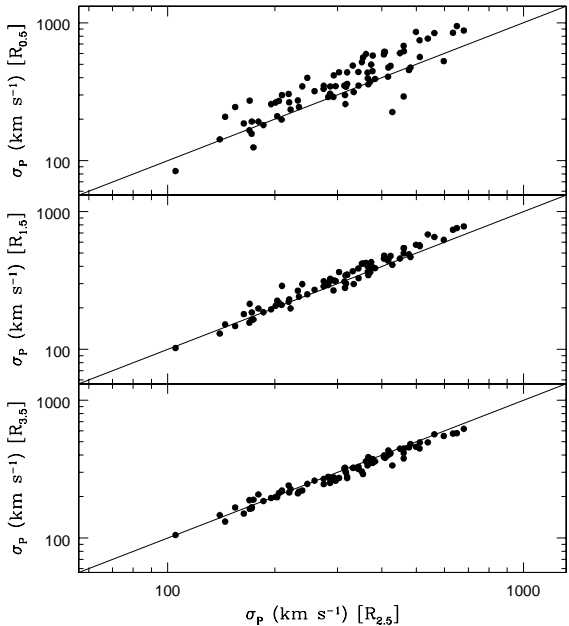}
\caption{Comparison of velocity dispersion estimates obtained for different maximum apertures to select galaxies from a spectroscopic survey around clusters. In the horizontal axis, we show the results considering R$_{max} = 2.5$ h$^{-1}$ Mpc. In the top panel the comparison is to the values of $\sigma_{P}$ with R$_{max} = 0.5$ h$^{-1}$ Mpc, while with R$_{max} = 1.5$ h$^{-1}$ Mpc and R$_{max} = 3.5$ h$^{-1}$ Mpc are shown in the middle and lower panel, respectively. The Y = X line is indicated in the three panels.} 
\label{fig:zspeczphot}
\end{figure}

\subsection{Luminosity limit}

Now we want to see how the magnitude limit of a given spectroscopic survey may affect the velocity dispersion estimates (as well as R$_{200}$ and M$_{200}$). From inspection of the magnitude and redshift distributions of the SDSS data, we find that the survey is approximately complete to $r = 17.6$ and $z = 0.1$, in agreement to the stated values ($r_{petro} = 17.77$). At $z = 0.1$ this magnitude limit allows one to sample M$^*_r + 1$ galaxies. At the bright-end of the luminosity function we expect the SDSS spectroscopic sample to be incomplete at $r < 14.5$ \citep {yor00}. 

To test the effect of missing galaxies at the bright or faint end of the luminosity function, we simply cut the SDSS data at different magnitude limits. For clarity, we show in Figure 8 only the results when considering galaxies brighter than $r = 16.6$. On every panel, the vertical axis has the results obtained with the full spectroscopic sample, while the horizontal axis has the results for galaxies brighter than r=16.6. The upper panel shows the comparison for the velocity dispersion, while the characteristic radius (R$_{200}$) and mass (M$_{200}$) are exhibited in the middle and lower panel, respectively. When restricting the sample to galaxies at $r \ge 14.5$, the velocity dispersion estimates are not affected by missing a few bright objects, even for the low redshift clusters. However, when selecting only galaxies at $r \le 16.6$ we notice that many clusters have their estimates affected by missing galaxies in the range $M^* < M < M^* + 1$. The filled circles in the figure represent clusters at $z \le 0.065$, which sample at least M$^*_r + 2$ with the full spectroscopic sample (complete to $r = 17.6$) and M$^*_r + 1$ for the limited sample (at $r = 16.6$). We see that for these clusters the scatter is smaller compared to the open circles, indicating that missing galaxies at $M^*+ 1 < M < M^* + 2$ have only a minor effect on the $\sigma_{P}$ estimates (as well as in the values of R$_{200}$ and M$_{200}$). 

Using the results shown in Figure 8 we can compute the mean and standard deviation of the difference between the velocity dispersion estimated with the full spectroscopic sample and only galaxies brighter than $r = 16.6$. We find $\mu = $ 3.09 $km$ $s^{-1}$ and $\sigma =$ 95.74 $km$ $s^{-1}$. In terms of mass, we have $\mu = $ 0.02 10$^{14}$ M$_{\odot}$ and $\sigma =$ 1.55 10$^{14}$ M$_{\odot}$. Although, the offset is consistent to zero we stress the large scatter. If we make a cut to consider only clusters with $\sigma_{P} \le$ 300.00 $km$ $s^{-1}$ (velocity dispersion of the X axis in Figure 8) we have the following: for velocity dispersion $\mu = $ 28.89 $km$ $s^{-1}$ and $\sigma =$ 85.11 $km$ $s^{-1}$, while for mass $\mu = $ 0.32 10$^{14}$ M$_{\odot}$ and $\sigma =$ 1.11 10$^{14}$ M$_{\odot}$. From that we see a small trend for underestimation of $\sigma_{P}$ and mass for low mass systems.

\begin{figure}
\begin{center}
\leavevmode
\includegraphics[height=0.50\textwidth,angle=0]{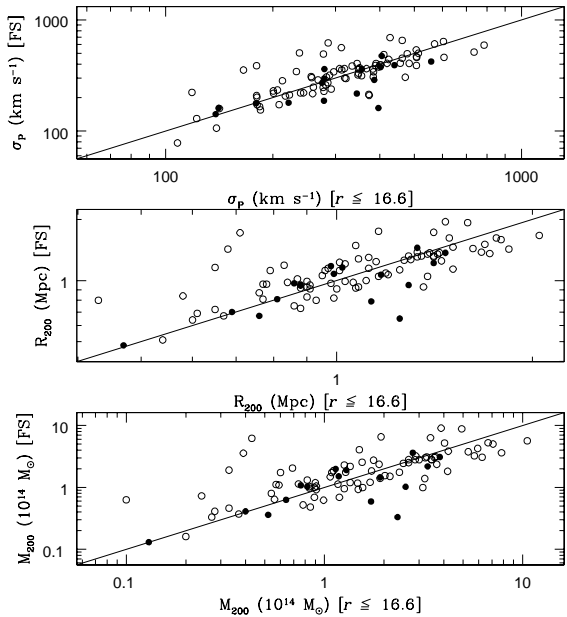}
\end{center}
\caption{Comparison of velocity dispersion, characteristic radius, and mass estimates obtained for different cuts in luminosity. The vertical axis in all panels has the results with the full spectroscopic sample, while  the horizontal axis has the results for galaxies brighter than $r = 16.6$. The filled circles represent clusters at $z \le 0.65$, which sample at least M$^*_r + 2$ with the full spectroscopic sample (complete to $r = 17.6$) and M$^*_r + 1$ for the limited sample (at $r = 16.6$). The Y = X line is indicated in all the panels.} 
\label{fig:zspeczphot}
\end{figure}

In the ideal case, we would like to have a spectroscopic survey as deep as possible. However, when that is not possible we can see from Figure 8 that a survey which is complete to $M^*+ 1$ is sufficient to get unbiased estimates of $\sigma_{P}$, R$_{200}$ and M$_{200}$. Figure 9 shows the phase-space diagram of the most massive NoSOCS cluster studied here, at $z = 0.0776$. In the upper panel we see all galaxies from the full spectroscopic survey, while the lower panel has the distribution of galaxies with $r \le 16.6$. We see that the absence of the $M^* < M < M^* + 1$ galaxies, in this case, {\it transforms} a massive cluster into a low richness system. The original estimates for this cluster are $\sigma_{P} = 693.57$ $km$ $s^{-1}$ and M$_{200} = 9.07$ 10$^{14}$ M$_{\odot}$. With the reduced data we find $\sigma_{P} = 427.28$ $km$ $s^{-1}$ and M$_{200} = 3.69$ 10$^{14}$ M$_{\odot}$. It is thus clear that using galaxy clusters at redshifts where the spectroscopic survey is very incomplete will severely affect the final mass estimates. 

\begin{figure}
\includegraphics[height=0.50\textwidth,angle=0]{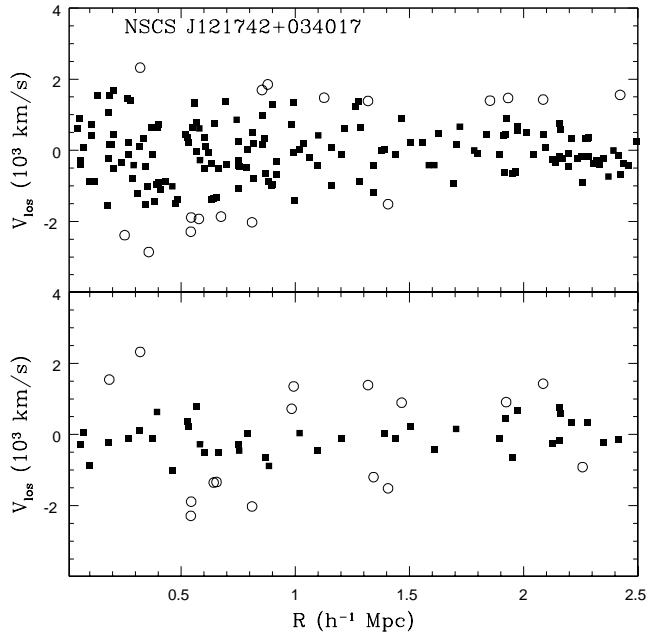}
\caption{Velocity diagram of the most massive NoSOCS cluster in the sample studied here. In the upper panel,  we show the original results obtained with the full spectroscopic survey of SDSS (same as in Figure 4). In the bottom panel we show only galaxies at $r \le 16.6$, to see how the interloper rejection, velocity dispersion, and mass estimates are affected. The original values for this cluster are $\sigma_{P} = 693.57$ $km$ $s^{-1}$ and M$_{200} = 9.07$ 10$^{14}$ M$_{\odot}$. Excluding galaxies at $r > 16.6$, we obtain $\sigma_{P} = 427.28$ $km$ $s^{-1}$ and M$_{200} = 3.69$ 10$^{14}$ M$_{\odot}$.}
\label{fig:zspeczphot}
\end{figure}

\subsection{X-ray luminosity from the ROSAT All Sky Survey (RASS)}

Following \citet {gal08} we used the locations of optically selected cluster candidates to measure X-ray fluxes and luminosities from RASS. The significance of X-ray emission in some of these areas may be too low to identify extended sources. However, data from this survey are still useful for deriving either fluxes or upper limits following \citet {boh00}. Although the X-ray data from RASS are limited, especially for the lower mass systems, the knowledge of the cluster positions allows us to estimate X-ray measurements for a large portion of our sample, comprised of 127 NoSOCS clusters plus 56 CIRS systems. Note that all CIRS objects had X-ray luminosities previously measured (obtained from different X-ray catalogs). However, we decided to re-estimate those in order to have a homogeneous set of estimates.

The X-ray luminosities L$_X$ are estimated from count rates in ROSAT PSPC images. Images and exposure maps are retrieved from the ROSAT archive via FTP. We begin by estimating the background contribution. That is done in three different ways. The first estimate is identical to the procedure adopted for the Northern ROSAT All-Sky Galaxy Cluster Survey (NORAS, \citealt {boh00}), where the background is determined from a ring area centered on the source. The inner ring radius is 21$'$ and the outer is 41.3$'$. The annulus area is further divided in 12 sectors, and the background is estimated from the median counts of all sectors. Further details can be found in \citet {boh00}. The second background estimate is based on the selection of a hundred boxes of one-hundred square arc minutes over the RASS field of 6.5$^{\circ} \times$ 6.5$^{\circ}$ where the cluster is. In both cases, the median and rms (bkg$_{med}$ and bkg$_{rms}$) are taken (from the 12 sectors or the 100 boxes). The third approach is based on the counts obtained in this whole frame of 6.5$^{\circ} \times$ 6.5$^{\circ}$. For this work we estimate the flux for each cluster in three different apertures, 0.5 h$^{-1}$ Mpc, R$_{500}$ and R$_{200}$. Only clusters whose total counts are 3$\sigma$ above the background are considered reliable. All others are reported as upper limits.

To convert the measured total count rate into an unabsorbed X-ray flux in the nominal ROSAT energy band (0.1-2.4 keV) we use the PIMMS tool available through NASA HEASARC. We assume a Raymond-Smith (RS) spectrum \citep {rs77} to represent the hot plasma present in the intracluster medium, with a metallicity of 0.4 of the solar value (the results are insensitive to small variations in metallicity). The interstellar hydrogen column density along the line-of-sight is taken as the weighted average value of \citet {dic90} (that is selected using the ``X-Ray Background Tool'', also available through NASA HEASARC). The flux in the nominal ROSAT energy band is calculated assuming a fixed temperature of 5 keV, which is typical for clusters \citep {mar98}. The resulting luminosity is termed L$_{X5}$. Then we use an iterative procedure relying on the L$_X$-T$_X$ relation from \citet {mar98}. From L$_{X5}$ we find a new temperature  which is used to recalculate the luminosity based on an RS spectrum. The procedure is iterated until convergence is reached (when the change in temperature is T$_X <$ 1keV , comparable to the scatter in the L$_X$-T$_X$ relation). The procedure typically converges in two or three iterations. For both luminosity measures, we apply a k$-$correction \citep {boh00} to derive the X-ray luminosity in the rest frame 0.1 - 2.4 keV band. 

Figure 10 shows a comparison of the X-ray luminosities obtained for the three
different background estimates discussed above. As we can see the luminosities
obtained with the ``annulus'' tend to be lower than the two other cases for
the most poor systems. However, for systems with L$_{X}^{R200} > 0.1-0.2$ $10^{44}$ erg/s the ``annulus'' and ``boxes'' are consistent to each other. When the comparison regards the ``frame'' 
the luminosities with the ``annulus'' background are always smaller. The 
results with the ``frame'' are also generally higher than those based on the
background obtained with the ``boxes''. So, we conclude that the ``frame'' background generally provides luminosities that are overestimated compared to the other two backgrounds. The good agreement between ``annulus'' and ``boxes'' is also seen in Figure 11, where we show the ratio of luminosities as a function of R$_{200}$, N$_{gals}^{R200}$ and $\sigma_P$. It is clear that for the richer systems (N$_{gals}^{R200} > 40$) the X-ray luminosities are more reliable.

\begin{figure}
\begin{center}
\leavevmode
\includegraphics[width=3.5in]{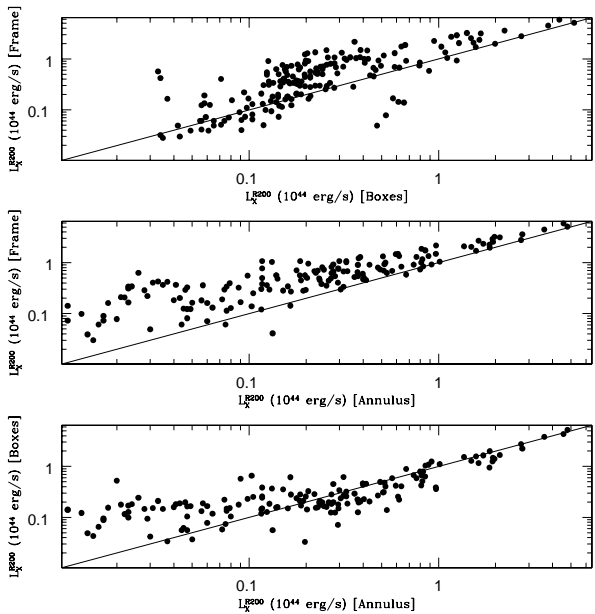}
\end{center}
\caption{Comparison of X-ray luminosities obtained for three different 
background estimates. In the lower panel the results considering the ``boxes''
background is compared to the ``annulus'', in the central panel the comparison
is between the ``frame'' and ``annulus''. In the upper panel luminosities
obtained with the ``frame'' are compared to the ``boxes''.}
\label{fig:richscalerelations}
\end{figure}

\begin{figure}
\begin{center}
\leavevmode
\includegraphics[width=3.5in]{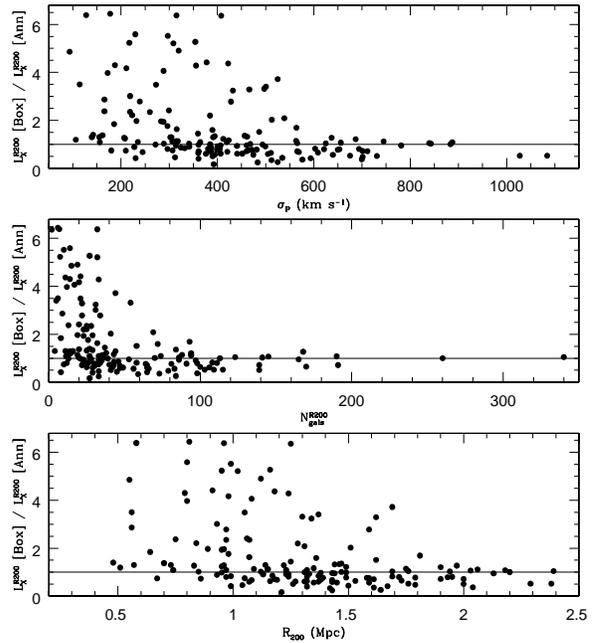}
\end{center}
\caption{Ratio of X-ray luminosities obtained with the  ``boxes'' and ``annulus'' backgrounds as a function of R$_{200}$ (lower panel), N$_{gals}^{R200}$ (middle) and $\sigma_P$ (upper panel).}
\label{fig:richscalerelations}
\end{figure}

Table 5 lists the values of L$_X$ (and its associated error) for the three
backgrounds; ``annulus'', ``frame'' and ``boxes'', respectively. We also list
the X-ray temperature measure from BAX (see section 4.5) when available and the interpolated temperature obtained from the L$_X$-T$_X$ relation. The last column of this table indicates whether we had a significant detection (SD), an upper limit (UL), or the source is too close to the border implying that the measurements might be taken with concern (XX). The last two columns are obtained with the background estimated from the annulus.

\begin{table*}
\begin{minipage}{160mm}
\caption{X-ray luminosity and temperature of the 183 NoSOCS plus CIRS clusters.}
\label{tab:cirs56}
\begin{tabular}{@{}lccccccc}
\hline
 name & L$_X$ (annulus) &  L$_X$ (frame) &  L$_X$ (box) & T$_X$ (BAX) & T$_X$ (interp. RASS) & NOTE \\
 & (10$^{44}$erg s$^{-1}$) & (10$^{44}$erg s$^{-1}$) & (10$^{44}$erg s$^{-1}$) & (keV) & (keV) & \\
\hline
NSCS J121847+484410 & 0.050 $\pm$  0.006 & 0.164 $\pm$  0.018 & 0.037 $\pm$  0.020 & ------ & 0.7 & SD \\
NSCS J011502+002441 & 0.404 $\pm$  0.019 & 0.508 $\pm$  0.042 & 0.229 $\pm$  0.049 & 2.53 $_{-0.09} ^{+0.09}$ & 2.1 & SD \\
NSCS J100242+324218 & 0.373 $\pm$  0.023 & 0.651 $\pm$  0.040 & 0.226 $\pm$  0.032 & ------ & 2.0 & SD \\

\hline
\end{tabular}
{Note. - A portion of this table is shown here for guidance regarding its form and content. A full version is available in the electronic edition of the MNRAS.}
\end{minipage}
\end{table*}

\subsection{Cluster masses estimated from T$_{X}$}

We searched BAX ({\it Base de Donn\'ees Amas de Galaxies X}, http://bax.ast.obs-mip.fr/) for counterparts of the 183 NoSOCS and CIRS clusters used in this work. We restricted the search to clusters at $z < 0.12$ with X-ray temperature measures available. We found 282 clusters in BAX, of which 21 are common to our sample. Actually, there were twenty-five clusters in common, but we rejected four that had no error estimates on their temperature. Data for these twenty-one systems come from  Einstein \citep {dav93}, BeppoSAX \citep {mar04}, ASCA \citep {fig01, ike02, fuz04}, XMM \citep {bel04} and Chandra \citep {mar00, yan04, vik06, fuj06}. The temperature values for these 21 clusters are listed in the fifth column of Table 5. To check the homogeneity of the temperature measures we selected from BAX, we show in Figure 12 the ratio between the temperature measures listed in \citet {rin06} and BAX. The former is retrieved from two sources only \citep {jon99, hor01}, while the latter comes from all the references listed above. As we see, the agreement for the 18 common clusters is very good, with almost all systems being concordant within 10\%.

For these clusters, we employed the M$_{200}$-T$_X$ relation given by equation 3 of \citet {pop05} to estimate mass from the temperature values. This relation was calibrated using a sample with more than one hundred clusters. Our goal is to compare the mass estimates obtained through different, independent methods. In one case the mass is estimated from the virial analysis applied to the galaxy distribution within clusters, while in the other case, mass is derived from the tight connection of the cluster mass to the temperature of the intra-cluster gas. Note that, from the 183 low$-z$ systems studied here, temperature is only available for generally massive clusters, with $\sigma >$ 400 km/s or M$_{200} >$ 10$^{14}$M$\odot$.

\begin{figure}
\begin{center}
\leavevmode
\includegraphics[width=3.5in]{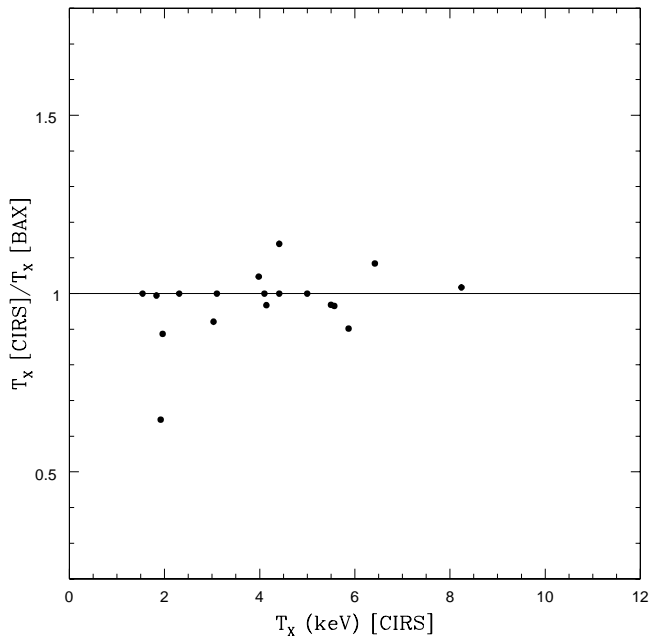}
\end{center}
\caption{Ratio between the temperature measures available in CIRS and selected from the literature (BAX). The results are shown as a function of the CIRS values. The figure shows the 18 common clusters with temperature available.}
\label{fig:richscalerelations}
\end{figure}

\section{Substructure estimates}

We have used photometric and spectroscopic data for galaxies in the NoSOCS and CIRS clusters to obtain an estimate of the fraction of systems with evidence for substructure. The codes used for the substructure analysis are those of \citet {pin96}, who evaluated the performance of thirty-one statistical tests. For the current work we consider the results of two of these tests, which are known to be the most sensitive. The first test considered is the DS or $\Delta$ test \citep {dre88}, which is a three dimensional test. The algorithm computes the mean velocity and standard deviation ($\sigma$) of each galaxy and its N$_{nn}$ nearest neighbors, where N$_{nn} = $ N$^{1/2}$ and N is the number of galaxies in the cluster region. The main difference to the approach of \citet{dre88} is that here we consider N$_{nn} = $ N$^{1/2}$, instead of N$_{nn} = 10$. Then these local means and $\sigma$ values are compared with the global mean and $\sigma$ (based on all galaxies). For every galaxy, a deviation from the global value is defined by the formula below. Substructure is estimated with the cumulative deviation $\Delta$ (defined by $\sum \delta_i$; see below). For objects with no substructure we have $\Delta \sim$ N. 

\bea
{\rm \delta_i^2} = \left( \frac{N_{nn}+1}{\sigma^2} \right)[(\bar{v}_{local}-\bar{v})^2 + (\sigma_{local} - \sigma)^2]. \,\,\,
\label{eq:perpdef}
\eea

The second test employed is two-dimensional, called symmetry or $\beta$ test and it was introduced by \citet {wes88}. The test looks for significant deviations from mirror symmetry about the cluster center. It assumes that a substructure represents a local asymmetry superposed on an otherwise symmetric distribution. For every galaxy ``i'', a local density estimate d$_i$ is obtained from the mean distance to the N$^{1/2}$ nearest neighbors (note that \citealt {wes88} used the five nearest neighbors). The local density, d$_{\circ}$, for a point ``$\circ$'' diametrical to a galaxy ``i'' is estimated in the same way. For a symmetric galaxy distribution the values of d$_i$ and d$_{\circ}$ should, on average, be approximately equal, but they will differ for clumpy distributions. The asymmetry for a given galaxy ``i'' is given by

\bea
{\rm \beta_i} =  \rmn{log}\biggl(\frac{d_{\circ}}{d_i}\biggr), \,\,\,
\label{eq:perpdef}
\eea

The $\beta$-statistic is then defined by the average value $<$$\beta_i$$>$ over all galaxies. For a symmetric distribution $<\beta> \approx 0$, while values of $<$$\beta$$>$ significantly greater than 0 indicate asymmetries. 

It is well known that any substructure test statistic has little meaning if not properly normalized, which can be achieved by comparing the results for the input data to those for substructure-free samples (the {\it null hypothesis}). For a two-dimensional test the null hypothesis is an azimuthally symetric smooth distribution, in which the surface density decreases with radius. For a three-dimensional test the null hypothesis is given by no correlation between position and velocity. In other words, velocity mean and dispersion should be the same locally and globally. For both tests, the significance level is determined through Monte Carlo simulations. For the $\beta$ test the null hypothesis files are created through azimuthal randomizations of the positions. In that case, the distance of each galaxy to the cluster center is kept, while an azimuth is randomly assigned. For the $\Delta$ test the velocities are shuffled randomly with respect to the positions, which remain fixed \citep{pin96}. 

For each input data set, we generate five-hundred simulated realizations. We then calculate the number of Monte Carlo simulations which show more substructure than the real data. Finally, this number is divided by the number of Monte Carlo simulations. For the current work we set our significance threshold at 5$\%$, meaning that only twenty-five simulated data sets can have substructure statistics higher than the observations to consider a substructure estimate significant. Further details can be found in \citet {pin96} and \citet{lop06}. The values of the two substructure tests and their significance levels are listed in the last four columns of Tables 1 and 3 for the NoSOCS and CIRS samples, respectively.

The substructure tests are only applied to clusters with at least five galaxies available within the aperture being used. Four different cases are considered. In the first we use the $\Delta$ test on the galaxies in the 170 clusters with at least five galaxies with available z$_{\rm spec}$ within R$_{200}$. We find that thirty-nine ($\sim$ 23\%) clusters show signs of substructure. In the second case, we applied the $\beta$ test to the same data set as above. This time, twenty-six ($\sim$ 15\%) are found to have substructure. Third, we applied the $\beta$ test to all galaxies within R$_{200}$ in each cluster with no restriction for using only galaxies that have z$_{\rm spec}$ available. The whole photometric data at m$^*-1 \le m_r \le $m$^*+1$ is used (see \citealt {lop06} for details regarding this choice). For this case, we find that forty-three ($\sim$ 24\%) clusters of all 179 show strong signs of substructure. Finally, we applied the $\beta$ test to all galaxies within 1.5 h$^{-1}$ Mpc of each cluster and found that sixty-four clusters ($\sim$ 35\%), of all 183, show strong signs of substructure. This last result is in line with the findings of \citet {lop06}. The rate of overlap between the four cases above is $\sim 50\%$.

\section{Comparison of optical and X-ray mass estimates}

In this section, we compare mass estimates obtained from the dynamical analysis of the galaxy distribution with the M$_{200}$-T$_X$ relation (T$_X$ given in BAX). In Figures 13 and 14, we show the ratio between the optical and X-ray masses. First, we considered the mass estimates obtained with the caustic technique \citep {rin06}. That is shown in Figure 13 for the nineteen CIRS clusters with measured gas temperatures. Then we show in Figure 14 the results achieved when considering the mass estimates from this work. In both plots, we show clusters with substructure (estimated with the $\Delta$ test) as open circles. The thin solid line on each figure shows the ratio equal to one, while the thick dashed line indicates the median value of the ratio for clusters with no substructure.

The inspection of these figures shows more clusters in Figure 14 agreeing within 40\% (the typical error in mass measurements) than in Figure 13. A few outliers displayed in Figure 14 (ratios lower than 0.5) are clusters affected by substructure. In this work, we find a good agreement with the X-ray expectations with a very weak trend in the sense of smaller X-ray to optical mass ratios for smaller masses. Moreover, we note that our result agrees to the one shown in Figure 1 of \citet {pop05}, where they also show a trend for optical estimates being slightly larger than the X-ray values.

In Figure 13 the two clusters with mass ratios close to 5 are Abell 1650 and MKW 08. We do not have a direct explanation for the lower value of M$_{200}$ found by the caustic technique. After searching for other temperature measures in the literature, we found those to be consistent to the ones we obtained from BAX. We can also see in Figure 14 that the virial mass we determined is well matched to the X-ray based mass, for these two clusters. The virial and X-ray masses differ only by 10.4\% for Abell 1650 and 2.6\% for MKW 08. The major outlier in Figure 14 is Abell 2197. The inspection of its phase-space diagram (Figure 5) does not qualify it as a problematic cluster. We note that this is one of the clusters found to have substructure (by both tests, $\Delta$ and $\beta$), which could explain the higher optical estimate. \citet {rin06} comment that A2197 is composed of two X-ray groups (A2197W and A2197E). The X-ray center is considered the peak of A2197W and the optical centroid lie within the two X-ray groups. 

We also notice that in Figure 13 most of the clusters with substructure (4 out of 6) have mass differences within 40\%, while in Figure 14 they all have differences higher than 40\%. From the comparison shown in Figure 14 we conclude that clusters with substructure generally have larger mass differences. That conclusion is not supported by Figure 13, indicating that the virial masses are more affected by projection or substructure, when compared to the caustic results. If we consider clusters with substructure to be not virialized the velocity dispersions generally overestimate the mass. The caustic mass is obtained directly from the mass profile, being independent of $\sigma_P$. The absolute median difference in Figure 13 is also higher than in Figure 14. However, the virial masses are generally higher than the X-ray masses. The opposite occurs when comparing X-ray and caustic masses. We also note there are more clusters closer to the ratio of one in Figure 14.

\begin{figure}
\begin{center}
\leavevmode
\includegraphics[width=3.5in]{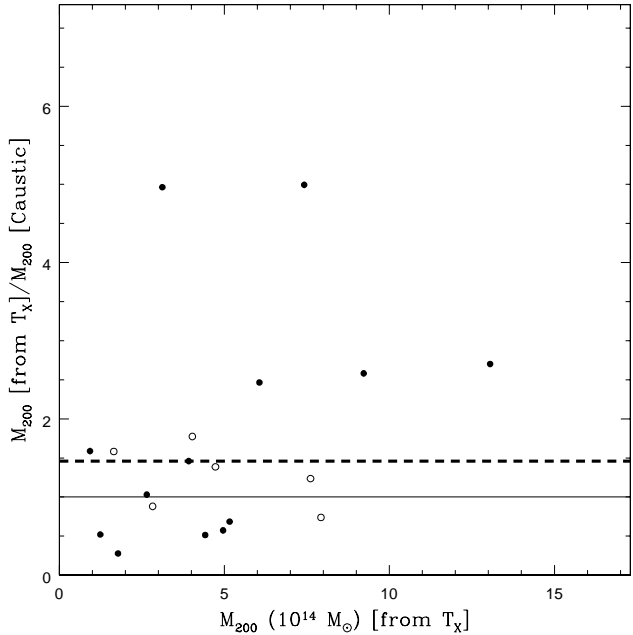}
\end{center}
\caption{The ratio between the mass obtained from the dynamical analysis of the optical data (masses measured by \citealt {rin06}) and the mass estimated from T$_{X}$ (selected from BAX). The figure shows the 19 common clusters with temperature available in BAX. Open circles represent clusters with substructure (estimated with the $\Delta$ test). The thin solid line represents the ratio of one, while the thick dashed line shows the median value for clusters without 
substructure (filled circles).}
\label{fig:richscalerelations}
\end{figure}

\begin{figure}
\begin{center}
\leavevmode
\includegraphics[width=3.5in]{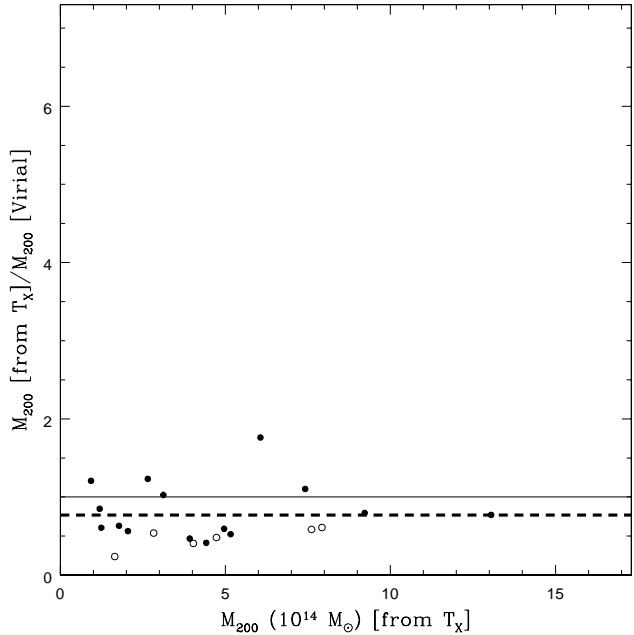}
\end{center}
\caption{The ratio between the mass obtained from the dynamical analysis of the optical data (masses from this work) and the mass estimated from T$_{X}$. The figure shows the 21 clusters with temperature available in BAX. Open circles represent clusters with substructure (estimated with the $\Delta$ test). The thin solid line represents the ratio of one, while the thick dashed line shows the median value for clusters without substructure (filled circles).}
\label{fig:richscalerelations}
\end{figure}

Considering the results presented in Figures 13 and 14 we carried out a more detailed comparison study with the results presented by \citet {rin06}. As both estimates are obtained from the galaxy distribution, we would not expect large differences. However, note that the interloper removal procedure is very different. Here, we apply the shifting gapper technique by \citet {fad96}, while \citet {rin06} employ the caustic method. The maximum aperture defining the region where galaxies are selected is also very distinct. \citet {rin06} consider an aperture of 10 h$^{-1}$ Mpc, four times the size considered here. Finally, the most relevant difference in the two approaches is that our masses result from the application of the virial theorem to the galaxy distribution, while \citet {rin06} estimate masses using the caustic technique (being R$_{200}$ and M$_{200}$ directly obtained from the caustic mass profile). 

In Figure 15, we show the comparison of the two optical mass estimates for the fifty-six clusters in common. Clearly many clusters disagree by more than 40\%. Substructure cannot explain the higher masses obtained by the virial analysis as there are clusters exhibiting substructure and still agree well in mass. We also noticed that clusters with higher ratios ($>$ 4) are generally the poorer ones and most of these have no indication of substructure. The higher masses obtained with the virial analysis are in agreement to what is seen in Figures 13 and 14, indicating a generally better agreement of our measurements to the X-ray estimates. The original CIRS values tend to be lower than the X-ray ones and thus tend to be lower than our results. \citet {rin06} argue that they found a good correlation between M$_{500}$ (caustic estimate) and T$_{X}$, as well as between M$_{200}$ (from the caustic) and $\sigma_P$ measured within R$_{200}$ (note that their $\sigma_P$ is independent of M$_{200}$ and R$_{200}$). They finally compare the caustic and virial masses at R$_{200}$ finding an excellent agreement between these measurements, thus concluding that the caustic mass estimate is unbiased. Note that \citet {rin06} consider the R$_{200}$ obtained by the caustic to determine the virial mass. 

The disagreement we find with \citet {rin06} could result primarily from the interloper removal procedure and the velocity dispersion estimates. However, we see in Figure 16 that just a few systems have estimates of $\sigma_P$ differing by more than 20\%. Hence, we conclude that the interloper removal method does not affect many clusters. Note also that our $\sigma_P$ values plotted are obtained only with cluster galaxies
within R$_{200}$. Those differ little from the ones obtained within the maximum aperture $R_A$ (generally close to 2.5 h$^{-1}$ Mpc) and listed in Table 4. However, that does not significantly affect the estimates if $R_A$ is sufficiently large (as seen in Figure 7). The good agreement exhibited in Figure 16 leads us to conclude that the difference in mass estimates does not result from the interloper removal method. 

Hence, we compare in Figure 17 the values of R$_{200}$. We see that our physical radius is generally different from the ones from the caustic. Namely, R$_{200}$ determined from the caustic method is systematically smaller than our values. Thus, the assumptions we make in the virial analysis lead to intrinsically different radius and mass estimates than the ones obtained from the mass profile, derived from the caustic analysis. The main reasons for the discrepancies may actually lie in the construction of the caustic mass profile, from which R$_{200}$ and M$_{200}$ are ultimately derived. Details about the caustic profile are found in \citet {rin06}. One important conclusion these authors reach is that the caustic mass profile is better represented by an NFW profile for 50\% of their sample. For 49\% of the clusters a better fit is given by a Hernquist profile and for 1\% by a singular isothermal sphere (SIS). In our approach we consider the NFW profile for representing the clusters. Depending of the concentration parameter the NFW profile may be lower than the caustic mass profile at low radius, resulting in a smaller R$_{200}$ value estimated by the caustic (see their Figure 12).

In summary, \citet {rin06} derive mass from the caustic mass profiles and they claim concordance between the caustic results and their virial masses. However, comparing our masses and the caustic values \citep {rin06} to the X-ray estimates, we find a better correlation with our virial masses. Therefore, we conclude that our estimates better represent the cluster potentials. That conclusion is also corroborated by the scaling relations we show in paper II \citep {lop08}, where we find more accurate results when using our mass estimates than the ones from \citet {rin06}. It is important to stress that \citet {rin06} use R$_{200}$ determined by the caustic to compute the virial masses. That choice contributes for the good agreement they find between the caustic and virial masses.

\begin{figure}
\begin{center}
\leavevmode
\includegraphics[width=3.5in]{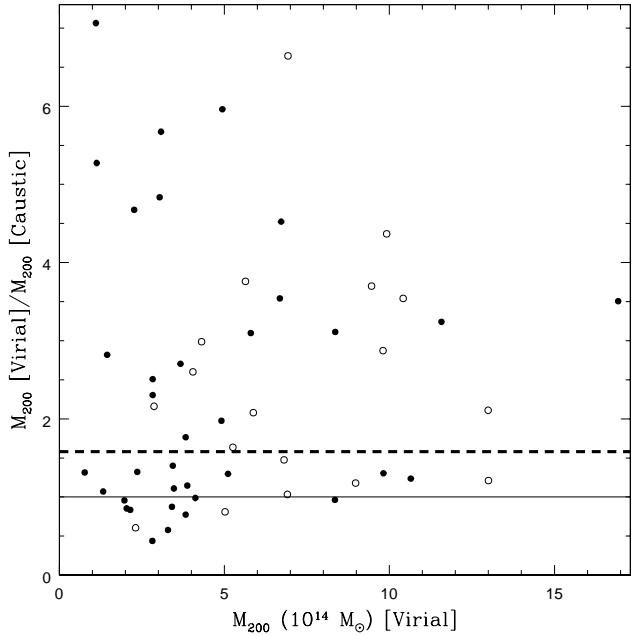}
\end{center}
\caption{Ratio between optical masses obtained by \citet{rin06} and in this work. The figure shows the 56 CIRS clusters used here. The thin solid line represents the ratio of one, while the thick dashed line shows the median value for clusters without substructure (filled circles).}
\label{fig:richscalerelations}
\end{figure}

\begin{figure}
\begin{center}
\leavevmode
\includegraphics[width=3.5in]{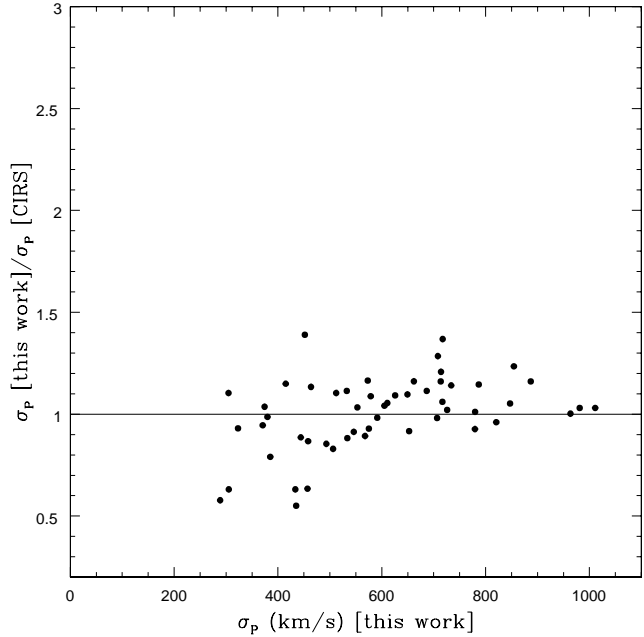}
\end{center}
\caption{Same as previous figure, but showing the results regarding the velocity dispersions.}
\label{fig:richscalerelations}
\end{figure}

\begin{figure}
\begin{center}
\leavevmode
\includegraphics[width=3.5in]{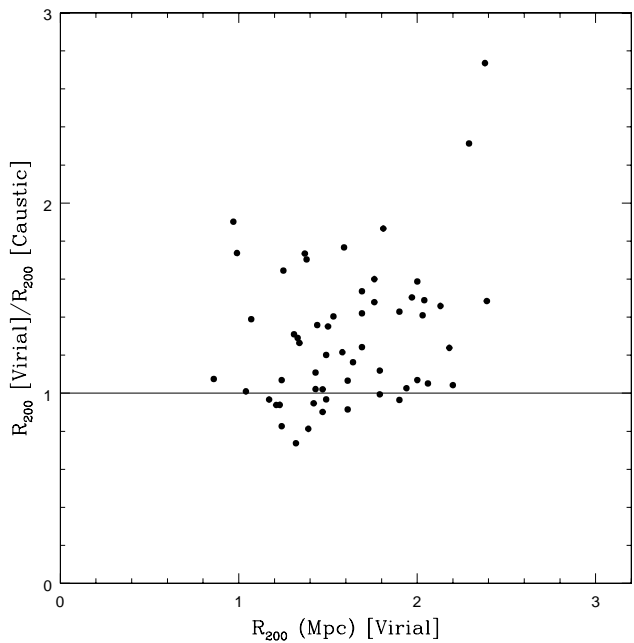}
\end{center}
\caption{Same as previous figure, but showing the results regarding the physical radius (R$_{200}$).}
\label{fig:richscalerelations}
\end{figure}

\section{Conclusions}

We have used SDSS data to study the galaxy properties of clusters first selected in the optical using DPOSS. We extracted SDSS data for every NoSOCS supplemental cluster \citep {lop03, lop04} sampled in DR5. Using the high quality photometric data of SDSS, we estimated new photometric redshifts (following \citealt {lop07}), richness and optical luminosity (\citealt {lop06}). After removing potential double clusters, we compiled a list with 7,414 systems ($\sim$ 75\% of the original NoSOCS supplemental sample) with properties derived from SDSS data. This cluster list has systems to $z \sim$ 0.5. When estimating richness, we also assess the effect of $k-correction$ and evolution in M$^*$ to the richness estimates.

For a subset of systems at $z \le 0.10$ we made use of the SDSS spectroscopic data to identify groups in redshift space, confirming these cluster candidates as true 3D overdensities. For every confirmed system, we applied the ``shifting gapper'' \citet {fad96} technique to exclude galaxy interlopers in the phase-space distributions and  estimate the velocity dispersions and masses (M$_{500}$ and M$_{200}$). The final sample with these properties derived at low-$z$ comprises 127 clusters. As the NoSOCS supplement does not sample very massive clusters at low$-z$, we extended the analysis for richer systems including those clusters from CIRS \citep {rin06}. Optical and dynamical properties are derived for fifty-six CIRS systems. Hence, the final list of clusters studied at low redshift ($z \le 0.10$) comprises 183 objects (127 from NoSOCS and 56 from CIRS). The main results of this work are:

\begin{enumerate}

\item We have checked how velocity dispersion and mass are affected by a less rigorous method for interloper removal, as well as the radial cutoff (maximum aperture sampled) and the luminosity limit of the spectroscopic survey. We found that looser criteria for removing interlopers can lead to higher mass estimates, which could bias the mass calibration (see discussion in paper II). With regards the maximum radius, we found that an aperture of 2.5 h$^{-1}$ Mpc is large enough to give unbiased velocity dispersion estimates. However, we notice that $\sigma_P$ and mass estimates are severely biased if one considers a spectroscopic survey which is not complete to at least M$^*+1$. That bias affects the construction of cluster scaling relations. 

\item Following \cite {gal08} we used RASS data to estimate X-ray luminosities for all 183 clusters. We obtained reliable X-ray detections for 85\% of the clusters, while for the remaining only upper limits were measured.

\item For a subset of these clusters (21 objects), we selected temperature measurements from the literature and estimated cluster mass assuming an M-T$_X$ relation. The comparison of optical and X-ray mass estimates for the subset of twenty-one clusters with measured T$_X$ reveals good concordance between masses from different wavelengths. We note a small trend for the optical masses estimated in this work to be larger than the X-ray values. However, we found the cluster masses determined from the caustic technique \citep {rin06} show a more pronounced trend for lower masses compared to masses determined from the gas temperatures. Thus, the optical masses from the caustic technique are generally smaller than the masses from the virial analysis obtained in the current work (which we conclude better represents the cluster potential). 

\item When comparing the velocity dispersion estimates of the present work to \citet {rin06} we do not see such a difference, which indicates that the interloper removal procedures lead to similar lists of cluster members. The comparison of R$_{200}$ values determined in this work and by \citet {rin06} reveals a systematic difference, which may explain the difference in mass. The mass discrepancies may be due to the shape of the caustic mass profile, which may result in biased values of R$_{200}$ and thus M$_{200}$. In \citet {lop08} we have also shown that the virial masses we obtained lead to more accurate scaling relations than the ones based on the caustic approach. More important, the scaling relations found with X-ray masses fully agree with the virial masses based relations, but show distinct results to the relations derived from the caustic masses. The scaling relations based in the caustic masses also show a much larger scatter than those based in the virial masses (see paper II; \citealt {lop08}).

\item For all 183 systems we have also estimated substructure from the galaxy distribution in two, and three dimensions, using the $\beta$ and $\Delta$ tests, respectively. We investigated the effect of substructure on the mass estimates. When considering only the spectroscopically confirmed cluster members, we find that 23\% of clusters have substructure estimated from the 3D $\Delta$ test, while 15\% are found to have substructure with the 2D $\beta$ test. When considering the photometric data set we find that 24\% of clusters have substructure according to the $\beta$ test within R$_{200}$. The same test applied to the photometric data within 1.5 h$^{-1}$ Mpc estimates in 35\% the ratio of systems with substructure. We also find that most of the clusters with substructure show large differences in mass estimates, when comparing X-ray and virial (this work) measures. The same is not true if the X-ray masses are compared to the caustic values. We conclude that substructure can partly explain the differences in the mass values from optical and X-ray data.

\end{enumerate}

\section*{Acknowledgments}

PAAL was supported by the Funda\c c\~ao de Amparo \`a Pesquisa do
Estado de S\~ao Paulo (FAPESP, processes 03/04110-3, 06/57027-4,
06/04955-1 and 07/04655-0). Part of this work was done at the 
Instituto Nacional de Astrof\'isica, Optica y Eletr\'onica and at the
Harvard-Smithsonian Center for Astrophysics. PAAL thanks the hospitality
during the stays in these two institutions. C. Jones thanks the Smithsonian 
Astrophysical Observatory for support.
The authors are thankful to A. Biviano for helpful discussions regarding
mass estimates of galaxy clusters, as well as for the suggestions made as the 
referee of the paper. These were very important for improving this paper.
We are thankful to Omar L\'opez-Cruz for discussions on richness estimates 
during the early stages of this work and to Jason Pinkney for making the 
substructure codes available. We would also like to thank Francesco 
La Barbera, Roy Gal and Marcelle Soares for fruitful discussions on 
this manuscript.

This research has made use of the NASA/IPAC Extragalactic Database
(NED) which is operated by the Jet Propulsion Laboratory, California
Institute of Technology, under contract with the National Aeronautics
and Space Administration.

Funding for the SDSS and SDSS-II has been provided by the Alfred
P. Sloan Foundation, the Participating Institutions, the National
Science Foundation, the U.S. Department of Energy, the National
Aeronautics and Space Administration, the Japanese Monbukagakusho, the
Max Planck Society, and the Higher Education Funding Council for
England. The SDSS Web Site is http://www.sdss.org/.

The SDSS is managed by the Astrophysical Research Consortium for the
Participating Institutions. The Participating Institutions are the
American Museum of Natural History, Astrophysical Institute Potsdam,
University of Basel, University of Cambridge, Case Western Reserve
University, University of Chicago, Drexel University, Fermilab, the
Institute for Advanced Study, the Japan Participation Group, Johns
Hopkins University, the Joint Institute for Nuclear Astrophysics, the
Kavli Institute for Particle Astrophysics and Cosmology, the Korean
Scientist Group, the Chinese Academy of Sciences (LAMOST), Los Alamos
National Laboratory, the Max-Planck-Institute for Astronomy (MPIA),
the Max-Planck-Institute for Astrophysics (MPA), New Mexico State
University, Ohio State University, University of Pittsburgh,
University of Portsmouth, Princeton University, the United States
Naval Observatory, and the University of Washington.




\label{lastpage}

\end{document}